\documentclass[amsmath,amssymb, aps,pra,twocolumn,superscriptaddress,nofootinbib,showkeys,a4paper]{revtex4-2}

\usepackage{orcidlink}
\usepackage{cancel}
\usepackage{datetime}
\usepackage{cancel}
\usepackage{braket}
\usepackage{booktabs}
\usepackage{framed}

\usepackage{graphicx}
\usepackage{siunitx}

\makeatletter
\newcommand*{\balancecolsandclearpage}{%
  \close@column@grid
  \cleardoublepage
  \twocolumngrid
}
\makeatother

\newcommand{\tvsw}[3]{\boldsymbol{\tilde #1}_{#2}^{(#3)}(k,\mathbf{r})}
\newcommand{\genbesselprime}{z_\ell^{(n)\prime}(kr)}
\newcommand{\tN}[2]{\tvsw{N}{#1}{#2}}
\newcommand{\tM}[2]{\tvsw{M}{#1}{#2}}
\newcommand{\tNlm}[1]{\tN{\ell m}{#1}}
\newcommand{\tMlm}[1]{\tM{\ell m}{#1}}
\newcommand{\genbesseli}[1]{z_\ell^{(#1)}(kr)}
\newcommand{\genbesselprimei}[1]{z_\ell^{(#1)\prime}(kr)}
\newcommand{\cN}[2]{\cvsw{N}{#1}{#2}}
\newcommand{\cNlm}[1]{\cN{\ell m}{#1}}
\newcommand{\cM}[2]{\cvsw{M}{#1}{#2}}
\newcommand{\cMlm}[1]{\cM{\ell m}{#1}}
\newcommand{\res}{\boldsymbol{\mathcal{R}}_n^{(\omega)}}
\newcommand{\lam}{\lambda}
\newcommand{\highlight}[2]{\colorbox{#1}{$\displaystyle#2$}}
\newcommand{\rp}{\rr_\mathbf{p}}
\newcommand{\mprime}{m^\prime}
\newcommand{\zsym}{\boldsymbol{\sigma}_z\!\,\text{-symmetry}}
\newcommand{\zz}{\phantom{+}0}

\newcommand{\cvsw}[3]{\boldsymbol{#1}_{#2}^{(#3)*}(k,\rr)}
\newcommand{\cPs}[2]{\cvsw{\Psi}{#1}{#2}}
\newcommand{\vsw}[3]{\boldsymbol{#1}_{#2}^{(#3)}(k,\rr)}
\newcommand{\uvsw}[3]{\mathbf{#1}_{#2}^{(#3)}(k,\rr)}
\newcommand{\cPlms}[1]{\cPs{\ell m\pol}{#1}}
\newcommand{\T}{\mathbf{T}}
\newcommand{\rr}{\mathbf{r}}

\newcommand{\N}[2]{\uvsw{N}{#1}{#2}}
\newcommand{\M}[2]{\uvsw{M}{#1}{#2}}
\newcommand{\Nlm}[1]{\N{\ell m}{#1}}
\newcommand{\Mlm}[1]{\M{\ell m}{#1}}
\newcommand{\Nlmn}{\N{\ell m}{n}}
\newcommand{\Mlmn}{\M{\ell m}{n}}
\newcommand{\genbessel}{z_\ell^{(n)}(kr)}
\newcommand{\Ps}[2]{\vsw{\Psi}{#1}{#2}}
\newcommand{\pol}{p}
\newcommand{\sphere}{{S_{R}^{2}}}
\newcommand{\dd}{\mathrm{d}}
\newcommand{\Plmns}{\Plms{n}}
\newcommand{\Plmnsprime}{\Ps{\ell^\prime m^\prime \pol^\prime}{n}}
\newcommand{\cPlmns}{\cPlms{n}}
\newcommand{\kron}[2]{\delta_{#1#2}}
\newcommand{\Plms}[1]{\Ps{\ell m\pol}{#1}}
\newcommand{\ii}{\mathrm{i}}
\newcommand{\gem}{\text{\char"00A4}}

\begin{document}\title{Pole-Expansion of the T-Matrix Based on a Matrix-Valued AAA-Algorithm}\author{Jan David Fischbach\orcidlink{0009-0003-8765-8920}
}\email{fischbach@kit.edu}
\affiliation{Institute for Nanotechnology, Karlsruhe Institute of Technology (KIT), Germany}\author{Fridtjof Betz\orcidlink{0000-0003-2193-7188}
}\affiliation{Zuse Institute Berlin, Germany}\author{Lukas Rebholz\orcidlink{0009-0001-0737-7826}
}\affiliation{Institute for Solid State Physics, Karlsruhe Institute of Technology (KIT), Germany}\author{Puneet Garg\orcidlink{0000-0002-7262-3085}
}\affiliation{Institute for Solid State Physics, Karlsruhe Institute of Technology (KIT), Germany}\author{Kristina Frizyuk\orcidlink{0000-0002-0506-464X}
}\affiliation{Institute for Solid State Physics, Karlsruhe Institute of Technology (KIT), Germany}\author{Felix Binkowski\orcidlink{0000-0002-4728-8887}
}\affiliation{Zuse Institute Berlin, Germany}\author{Sven Burger\orcidlink{0000-0002-3140-5380}
}\affiliation{Zuse Institute Berlin, Germany}\affiliation{JCMwave GmbH, Germany}\author{Martin Hammerschmidt\orcidlink{0000-0003-0291-1599}
}\affiliation{JCMwave GmbH, Germany}\author{Carsten Rockstuhl\orcidlink{0000-0002-5868-0526}
}\affiliation{Institute for Nanotechnology, Karlsruhe Institute of Technology (KIT), Germany}\affiliation{Institute for Solid State Physics, Karlsruhe Institute of Technology (KIT), Germany}\newdate{articleDate}{20}{2}{2026}
\date{\displaydate{articleDate}}
\begin{abstract}
The transition matrix (T-matrix) is a complete description of an object's linear scattering response. As such, it has found wide adoption for the theoretical and computational description of multiple-scattering phenomena. In its original form, the T-matrix describes the interaction of a scatterer with a monochromatic source. In practice, however, information about the T-matrix is usually needed in an extended spectral domain. To access the frequency-dispersion, one might naively sample T-matrices over a finely resolved set of discrete frequencies and store one T-matrix per frequency. This approach has multiple drawbacks: it is computationally expensive, requires excessive memory, and it disregards the physical origin of the spectral features, weakening physical interpretability. To overcome these major limitations, we leverage a pole-expansion technique to represent the T-matrix with arbitrary frequency resolution within a selected frequency domain via a set of resonant contributions. A matrix-valued variant of the recently established adaptive Antoulas-Anderson (AAA) algorithm for rational approximation enables us to compute the pole-expansion at minimal computational cost using only a small number of direct evaluations. We demonstrate the benefits of such a representation with examples ranging from semi-analytically accessible scatterers to quasi-dual bound states in the continuum. To allow the wider community to capitalize on these findings, we provide open-source tools to perform the presented pole-expansion of the T-matrix.
\end{abstract}
\maketitle

\section{Introduction}

Particles interacting with light at wavelengths comparable to their size are abundant in nature and in technology. To capture their response on theoretical grounds, Gustav Mie introduced his seminal scattering coefficients, characterizing individual spheres \citep{Mie_1908}. Their generalization to arbitrary scatterers -- the T-matrix \citep{waterman_matrix_1965} -- has since been widely employed in applications ranging from atmospheric sciences \citep{ito_t-matrix_2017}, to multiple-scattering metasurfaces for sensing \citep{patti_chiral_2019}, and solar applications \citep{jaeger_optics_2025}. The conventional approach to computing frequency-dispersive T-matrices relies on repeated evaluations over a finely resolved set of frequencies \citep{Asadova_2025}. While this method, in principle, provides an accurate description, it becomes computationally expensive when high spectral resolution is required, particularly when sharp resonant features dominate the response. Even more severe, in the interaction of multiple objects, each described by an individual T-matrix, sharp resonances may appear only from particle interactions \citep{kravetsPlasmonicSurfaceLattice2018}. In this case, it is even difficult to estimate the spectral range in which to search for resonances. As a result, the spectral response over a broad bandwidth must be sampled very finely to accurately capture those emerging features. To ease the situation, it is beneficial to represent the T-matrix $\T$ (or equivalently the scattering matrix $\mathbf{S}$) as a sum over resonant (pole-) contributions with analytically known frequency dependence. Two approaches can be distinguished to retrieve these poles.

%
%
%

\begin{figure*}[!htbp]
\centering
\includegraphics[width=0.8\linewidth]{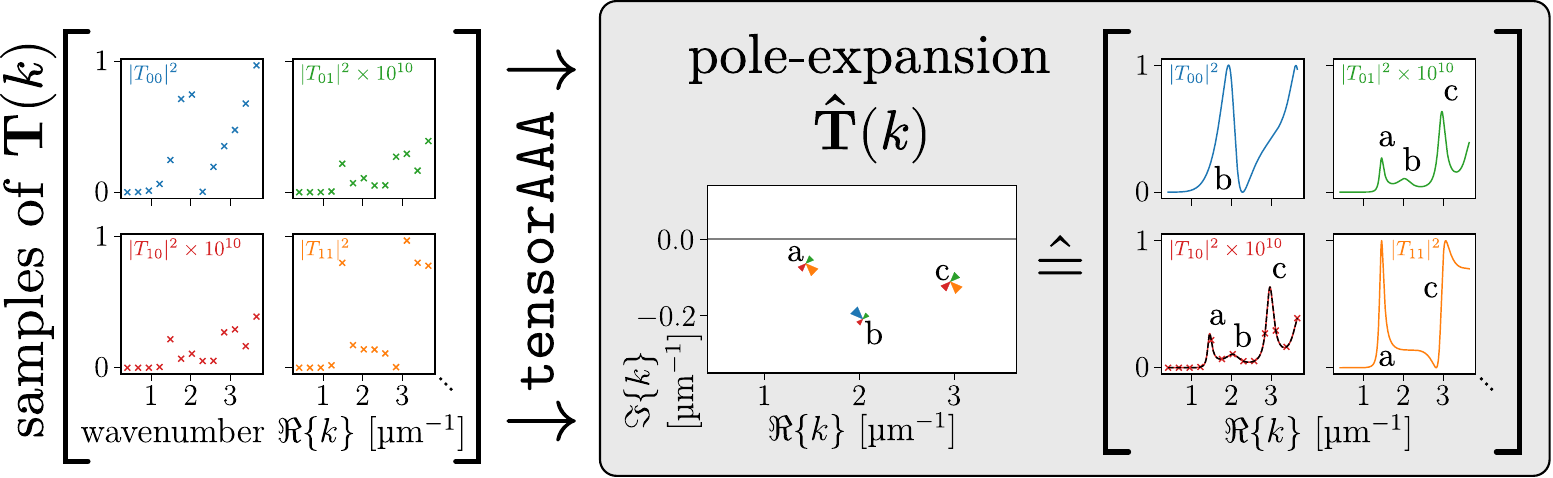}
\caption[]{Schematic representation of the introduced method to obtain a pole-expansion of the T-matrix. Samples of the T-matrix at different frequencies (the first four matrix elements $|T_{ij}|^2$ are indicated in the left panel) are jointly fed to a matrix-valued variant of the adaptive Antoulas-Anderson (AAA) algorithm (\texttt{tensorAAA}). The result is a rational approximation of the T-matrix, decomposing it into contributions of simple poles. The dominant poles coincide with the resonant states of the scatterer, while others approximate a slowly varying background term [see Eq. (\ref{eq-pole-exp-T})]. The central panel shows the dominant poles in the complex plane, schematically indicating the matrix valued residues by the size of four triangle markers corresponding to the four entries of the T-matrix shown in the other panels. The three poles are labeled a,b, and c. As shown in the right panel, this representation can be evaluated for arbitrary frequencies practically for free. The poles correspond to peaks (same labels) in the spectrum, where the width is determined by the pole's imaginary part and the intensity is jointly determined by the residue and imaginary part. To emphasize the remarkable efficiency regarding the number of required sample points for the \texttt{tensorAAA}, the samples from the left panel are overlayed for the matrix elements colored in red. To highlight the accuracy, reference solutions with fine frequency resolution are shown as a black dashed line for the same matrix element. This figure is meant purely as a high-level conceptual introduction of our method. The treated scatterer is a sphere with a chiral shell mixing the electric and magnetic scattering coefficients. Further details can be found in Supplement I (`Sphere with Chiral Shell').}
\label{fig-intro}
\end{figure*}

In a first approach, a large body of work has developed pole-expansions by computing the resonant states of a given scatterer, \textit{i.e.}, by solving the nonlinear eigenvalue problem posed by the sourceless Maxwell equations with outgoing boundary conditions \citep{bothResonantStatesTheir2022, weiss_how_2018, wu_intrinsic_2020}. However, to accurately recover the scattering behavior, this approach typically requires a large number of modes [sometimes including modes introduced by the discretization of the outgoing boundary conditions, known as perfectly matched layer (PML) modes] \citep{sauvanNormalizationOrthogonalityCompleteness2022}. Alternatively, the remaining error can be encapsulated into a background term found by interpolating a coarser set of frequency samples \citep{weiss_how_2018, wu_efficient_2021}.

The second approach, which we take here, extracts a pole-based representation directly from such a sampled response, avoiding the explicit treatment of the nonlinear eigenproblem entirely. The adaptive Antoulas-Anderson (AAA) algorithm \citep{nakatsukasaAAAAlgorithmRational2018} has recently been proposed as a suitable tool to obtain accurate representations of \textit{scalar} optical transfer functions \footnote{Here we follow the convention of Ref. \citep{SalehAppendixLinearSystems1991}, using the term \textit{(optical) transfer function} for the frequency domain description of the systems response to some excitation. Ref. \citep{SalehAppendixLinearSystems1991} uses the term \textit{impulse response function} to refer to the time domain response to the excitation by a Dirac-delta impulse. The response to an arbitrary excitation is found as the convolution of the excitation and the \textit{impulse response function}. The \textit{transfer function} results from the \textit{impulse response function} via the Fourier-transform.} in the form of pole-exansions \citep{betzEfficientRationalApproximation2024a}. A comprehensive description of a given scatterer requires one to consider the response under different incident fields. The transition from multiple input channels to multiple output channels naturally leads to a matrix-valued transfer function. In this contribution we will focus on the frequency dependent T-matrix $\T(k)$, which maps the incident field coefficients to scattered field coefficients in a basis of vector spherical waves (VSWs) \citep{beutelTreamsTmatrixbasedScattering2024}. For consistency with existing multiple-scattering literature, we express the frequency dependence in terms of the wavenumber $k$. Simply applying the scalar AAA-algorithm to each matrix element individually has the following drawbacks:

\begin{enumerate}
\item The poles obtained for different matrix elements differ, preventing a consistent physical interpretation of the poles in terms of resonances.
\item Every matrix element requires storing and evaluating its own representation, leading to excessive computational and storage cost.
\end{enumerate}

In this contribution, we show that a matrix-valued variant of the AAA-algorithm enables the simultaneous pole-expansion of \textit{all} entries of the T-matrix using a joint set of poles (see Figure~\ref{fig-intro}). We will illustrate the convergence of such a pole-expansion of $\T$ and demonstrate its benefits for applications that require a high frequency resolution. This allows us to recover the radiative multipolar contents of the corresponding resonant states from the residues of the expansion. As such, our method enables us to straightforwardly uncover the multipolar nature of multiple-scattering resonances.

\section{Theoretical Background}

In this section, we will provide the theoretical foundation, briefly reviewing the propagation of electromagnetic fields in a homogeneous embedding and the T-matrix formalism based on vector spherical waves to describe arrangements of scatterers embedded in such media. Further, we outline the pole-expansion technique for scalar transfer functions, which we will use as the basis to later introduce its matrix-valued extension, obtaining the pole-expansion of the T-matrix.

\subsection{Electromagnetic Fields in a Homogeneous Embedding}

The Maxwell equations for time harmonic fields in isotropic homogenous space read as \citep{beutelTreamsTmatrixbasedScattering2024}:
\begin{equation}
\nabla\times
\begin{pmatrix}
\mathbf{E}(k, \rr) \\
Z_0\mathbf{H}(k, \rr)
\end{pmatrix}=k
\begin{pmatrix}
0 & \mathrm{i}\mu(k) \\
-\mathrm{i}\varepsilon(k) & 0
\end{pmatrix}
\begin{pmatrix}
\mathbf{E}(k, \rr) \\
Z_0\mathbf{H}(k, \rr)
\end{pmatrix}.
\end{equation}
Here, the frequency dependence is represented in terms of the wavenumber $k = \omega/c_0 = \omega\sqrt{\varepsilon_0\mu_0}$. Further, we use the $e^{ -\mathrm{i} kc_0 t}$ convention for time harmonic fields.
$\mathbf{E}(k, \rr)$ and $\mathbf{H}(k, \rr)$ are the electric and magnetic fields, respectively. Further, $\varepsilon(k)$ and $\mu(k)$ are the dispersive permittivity and permeability tensors, which encapsulate the response of the homogeneous background material in terms of transfer functions. In vacuum, they correspond to $\varepsilon_0$ and $\mu_0$. Both quantities define the vacuum impedance $Z_0 = \sqrt{\frac{\mu_0}{\varepsilon_0}}$ and the speed of light in vacuum $c_0 = 1/\sqrt{\varepsilon_0\mu_0}$, respectively.

Because Maxwell's equations are linear, their solutions can be represented as a superposition of basis solutions. One such basis set, particularly suitable for the description of light scattered from finite-sized objects, is the set of vector spherical waves \citep{beutelTreamsTmatrixbasedScattering2024}:

\begin{equation}
\begin{aligned}
\Mlmn &= \nabla \times [\rr \genbessel Y_\ell^m(\theta, \varphi)] \, ,
\\
\Nlmn &= \frac{\nabla}{k} \times \Mlmn.
\end{aligned}
\end{equation}
Here, $(r, \theta, \varphi)$ express $\mathbf{r}$ in polar coordinates and $Y_\ell^m(\theta, \varphi)$ are the scalar spherical harmonics \citep{jackson_classical_1999}. $\genbessel$ are the spherical Bessel or Hankel functions of first or second kind depending on the superscript $n$, following the convention of references \citep{beutelTreamsTmatrixbasedScattering2024, santiago_decomposition_2019}. To streamline the notation, we define
\begin{equation}
\mathbf{\Psi}_{\ell m\pol}^{(n)}(k, \rr) = \begin{cases}
      \Nlmn, & \text{if}\ \pol=1 \\
      \Mlmn, & \text{if}\ \pol=-1
    \end{cases}\quad.
\end{equation}
For $k\in\mathbb{R}$, the VSWs fulfill the following orthogonality relation when integrating over the sphere $\sphere$ with radius $R$ \citep{santiago_decomposition_2019}:

\begin{equation}
\label{eq:ortho}
\frac{\int_\sphere \cPlmns \cdot \Plmnsprime \, \dd S}{\int_\sphere \left|\Plmns \right|^{2}\dd S}=\kron{\ell^\prime}{\ell}\kron{m^\prime}{m}\kron{\pol^\prime}{\pol} \, .
\end{equation}
The orthogonality of VSWs enables the decomposition of the field incident on a scatterer in terms of the set of regular VSWs $\Plms{1}$ \citep{beutelTreamsTmatrixbasedScattering2024}:
\begin{equation}
\begin{aligned}
    \mathbf{E}_\mathrm{inc}(k, \rr)=\sum_{\ell=1}^{\infty}\sum_{m=-\ell}^{\ell}\sum_\pol a_{\ell m \pol}(k)\Plms{1} \, .
\end{aligned}
\end{equation}
A similar decomposition is possible for any field scattered by an object outside its circumscribed sphere in terms of outgoing VSWs $\Plms{3}$:

\begin{equation}
\begin{aligned}
    \mathbf{E}_\mathrm{sca}(k, \rr)=\sum_{\ell=1}^{\infty}\sum_{m=-\ell}^{\ell} \sum_\pol q_{\ell m\pol}(k)\Plms{3}\, .
\end{aligned}
\end{equation}

\subsection{The T-Matrix}\label{sec:tmat}

By collecting the coefficients of the incident and scattered fields in the (infinite-dimensional) vectors $\mathsf{a}(k)$ and $\mathsf{q}(k)$, respectively, we can express the linear scattering response for any incident field by the infinite-dimensional operator $\mathsf{T}(k)$ \citep{beutelTreamsTmatrixbasedScattering2024}:
\begin{equation}
\mathsf{q}(k) = \mathsf{T}(k) \mathsf{a}(k)\, .
\end{equation}
For practical purposes, the $\mathsf{T}(k)$ operator is typically truncated to a matrix of finite size $\mathbf{T}(k)$ (\textit{e.g.}, by choosing a maximum multipolar degree $\ell_\mathrm{max}$) and sampled at discrete frequencies. Here, we denote the corresponding truncated coefficient vectors as $\mathbf{a}(k)$ and $\mathbf{q}(k)$, respectively. Such a truncation enables the numerical treatment of various multiple-scattering phenomena. Applications include scattering in disordered media \citep{theobald_simulation_2021}, regular lattices of molecules \citep{zerulla_multi-scale_2022} or other scatterers \citep{beutel_efficient_2021}, and clusters of particles, such as plasmonic aggregates \citep{dwivedi_effective-medium_2024}, where interactions by multiple scatterers play a crucial role in shaping the overall response. With appropriate modifications, the T-matrix description extends to scatterers on substrates \citep{egel_celes_2017} and even relativistically moving objects \citep{whittam_circular_2024}.

The particular beauty of a T-matrix-based description of a scatterer lies in the possibility to compute the response of arrangements of known scatterers when these are radiatively coupled. Leveraging the knowledge of geometric series, the Born series of repeated scattering events between an individual scatterer and all others can be encapsulated into a single matrix solve \citep{beutelTreamsTmatrixbasedScattering2024}:
\begin{equation}
\label{eq:cluster_tmat}
\mathbf{q}_\mathrm{local}(k) = \left[\mathbf{1} - \mathbf{T}_\mathrm{diag}(k) \mathbf{C}^{(3)}(k)\right]^{-1} \mathbf{T}_\mathrm{diag}(k) \mathbf{a}_\mathrm{local}(k) \, .
\end{equation}
Here, $\mathbf{T}_\mathrm{diag}(k)$ is a block-diagonal matrix containing the individual T-matrices of all scatterers, while $\mathbf{C}^{(3)}(k)$ encapsulates the translations between scatterers and the basis change from outgoing VSWs $\Plms{3}$ to regular VSWs $\Plms{1}$ \citep{waterman_matrix_1965, beutelTreamsTmatrixbasedScattering2024}. Further, $\mathbf{a}_\mathrm{local}$ and $\mathbf{q}_\mathrm{local}$ are the incident and scattered field coefficients in the local basis of the individual scatterers \citep{beutelTreamsTmatrixbasedScattering2024, suryadharmaStudyingPlasmonicResonance2017}. As Eq. (\ref{eq:cluster_tmat}) describes the scattered field under arbitrary illuminations, it permits the extraction of a T-matrix describing the cluster as a single compound scatterer \citep{beutelTreamsTmatrixbasedScattering2024}.
When scatterers are placed on a regular lattice to form a metasurface and illuminated with a fixed parallel component of the wave vector $\mathbf{k}_\parallel$, the lattice T-matrix $\mathbf{T_\mathbf{k_\parallel}}(k)$ (sometimes denoted as renormalized T-matrix) can be similarly obtained by replacing $\mathbf{C}^{(3)}(k)$ with a matrix that sums over the contribution of all lattice sites. Ewald summation techniques enable the efficient evaluation of such lattice sums \citep{beutel_efficient_2021}. The lattice T-matrix enables calculating the scattered field from a metasurface by expanding the incident field into periodically repeated regular VSWs and applying $\mathbf{T_\mathbf{k_\parallel}}(k)$ once to obtain the scattered field in a basis of periodically repeated radiating VSWs. These periodically repeated radiating VSWs interfere constructively only in the permitted diffraction orders, yielding the response in terms of radiating and evanescent plane waves \citep{liuBeamSteeringDielectric2018}.




Various computational methods have been developed to obtain the underlying T-matrices. For some highly symmetric scatterers, semi-analytic expressions exist \citep{mie_beitrage_1908}, whereas in the general case, numerical methods are necessary to determine the matrix elements of $\T(k)$. Approaches to computing the T-matrix can be broadly categorized by their use of interface conditions. The extended boundary condition method (EBCM) \citep{waterman_matrix_1965} and leverage an interface-based description to obtain the T-matrix. Extraction from finite-difference time-domain (FDTD) \citep{oskooi_meep_2010} simulations, using the discrete dipole approximation (DDA) \citep{draine_discrete-dipole_1994, yurkin_discrete-dipole-approximation_2011}, and from finite element methods (FEM) \citep{santiago_decomposition_2019} rely on illuminating the scatterer with a set of different incident waves -- commonly plane \citep{fruhnert_computing_2017} or spherical waves -- to reconstruct the T-matrix. A common approach involves illuminating the scatterer with individual basis vectors of the incident field (\textit{i.e.}, the regular VSWs). The scattered field is then decomposed into its basis vectors (\textit{i.e.}, outgoing VSWs) yielding entire columns of the T-matrix at a time. Using the orthogonality relations [Eq. (\ref{eq:ortho})], the matrix elements are computed as \citep{santiago_decomposition_2019}:

\begin{equation}
\label{eq:decomposition}
T_{\ell m \pol, \ell^\prime m^\prime \pol ^\prime} (k) =\frac{\int_{\sphere}\mathbf{\Psi}_{\ell m \pol}^{(3)*}(k, \rr)\cdot\mathbf{E}_{\mathrm{sca}, \ell^\prime m^\prime \pol^\prime}(k, \rr) \, \dd S}{\int_{\sphere}\left|\mathbf{\Psi}_{\ell m \pol}^{(3)}(k, \rr)\right|^2 \dd S}
\end{equation}
Where $\mathbf{E}_{\mathrm{sca}, \ell^\prime m^\prime \pol^\prime}(k, \rr)$ denotes the scattered field under illumination by the regular VSW $\mathbf{\Psi}_{\ell m\pol}^{(1)}(k, \rr)$. A more comprehensive overview of different methods to determine $\T(k)$ is provided in \citep{Asadova_2025}.

\subsection{Pole-Expansion of a Single T-Matrix Element}

It is inherent to Maxwell's equations, that the same scatterer acts differently when exposed to different frequencies of light, which is expressed here by the $k$-dependence of the T-matrix $\T(k)$. Nonetheless, the response at different $k$ is not independent. Similar to the Kramers-Kronig relations for material transfer functions \citep{landau_electrodynamics_1984}, the linearity and causality of Maxwell's equations impose powerful analytic constraints on the $k$-dependence of a wide variety of scalar optical transfer functions. Typical examples of such scalar transfer functions are the normalized decay rate of a dipole emitter (Purcell enhancement \citep{Purcell1946}), the scattered field amplitude at a fixed point in space, or a far-field amplitude, just to name a few. Such transfer functions can be analytically continued into the complex plane. However, it is not possible to find a continuation that is analytic on the entire complex plane, but rather isolated singularities and other special locations (\textit{i.e.}, branch cuts and accumulation points) have to be excluded. Isolated singularities (which are typically of the first order) coincide with the poles of the Green's dyadic \citep{binkowski_poles_2024}, which are thoroughly discussed in literature, where they appear under various names like resonant states \citep{bothResonantStatesTheir2022, muljarov_resonant-state_2016}, quasi-normal modes (QNMs) \citep{sauvanNormalizationOrthogonalityCompleteness2022, kristensen_modes_2014, wuNanoscaleLightConfinement2021}, scattering modes \citep{morse_methods_1946}, resonances, or simply modes \citep{binkowski_poles_2024}. Expanding the quantity of interest into resonant contributions and a slowly varying background $\mathrm{BG}(k)$ yields a convenient and physically meaningful representation, which we call pole-expansion \citep{binkowski_poles_2024}:
\begin{equation}
\label{eq:pole_expansion}
f(k) \approx r(k) = \sum_n \frac{\mathcal{R}_n}{k-k_n}  + \mathrm{BG}(k)\, .
\end{equation}
Here, $f(k)$ is the quantity of interest, $r(k)$ is the approximant, and $(k_n, \mathcal{R}_n)$ are the pole-residue pairs. The residues $\mathcal{R}_n$ depend on the variable in which the pole-expansion is performed. For expansions in angular frequency $\omega=c_0 k$ the residues relate to those in wavenumber $k$ as $\mathcal{R}_n^{(\omega)} = \mathcal{R}_n^{(k)} / c_0$.

From the sampled response, the framework of complex analysis allows one to find the poles and residues in several ways, \textit{e.g.}, by contour integration \citep{zolla_foundations_2012, zschiedrich_riesz-projection-based_2018} or techniques for rational approximation like vector fitting \citep{gustavsen_rational_1999}, the Loewner framework \citep{antoulas_scalar_1986}, or the adaptive Antoulas Anderson (AAA) algorithm \citep{nakatsukasaAAAAlgorithmRational2018, betzEfficientRationalApproximation2024a}. Some of these approaches require evaluating the response (or better, its analytic continuation) in the complex plane. In Supplement II (`Analytic Continuation of the VSW Decomposition Integrals'), we provide the analytic continuation to the surface integral, decomposing the scattered field into VSW contributions [Eq. (\ref{eq:decomposition})]. Consequently, the scalar matrix elements of the T-matrix $f(k) = T_{ij}(k)$ can be expanded into pole contributions using any of the methods above.

Our method of choice is the AAA-algorithm, which has proven to be robust and versatile in many applications scenarios \citep{nakatsukasaApplicationsAAARational2025}. At the heart of the AAA-algorithm is the barycentric form
\begin{equation}
r(k) =  \frac{\sum_j \frac{w_j f_j}{k-k_j}}{\sum_j \frac{w_j}{k-k_j}}
\end{equation}
that is iteratively built from a subset of the samples ($k_j$, $f_j=f(k_j)$), which are interpolated by construction. After each iteration the weights $w_j$ are adjusted to minimize the error with respect to the remaining samples in a least-square sense. The iteration is stopped when this error meets a user-defined tolerance or a maximum number of iterations is reached.

Poles and residues can be directly obtained from the barycentric rational form. It is common practice to consider poles within a domain of interest, in which the approximation is accurate, as the system's resonances. In contrast, poles outside this domain of interest are attributed to the background $\mathrm{BG}(k)$. When constructing the approximation exclusively from samples on the real frequency axis, it is not clear how far the domain of validity reaches into the complex plane. Augmenting the AAA-algorithm with an iterative sample-refinement strategy that adaptively adds samples in the complex plane near the identified resonances allows one to define the domain more clearly \citep{fischbachFrameworkComputeResonances2025}.

\section{Pole-Expansion of the T-matrix}

Above, we presented an established method to construct a pole-expansion of a scalar optical transfer function, and how it can be applied to individual matrix elements $T_{ij}(k)$. However, it is desirable to find one set of poles for all entries of $\T(k)$ simultaneously. Otherwise, slightly shifted poles in each element will obfuscate their joint origin, which is rooted in the physical reality of one resonant state coupling to multiple radiation channels. Moreover, a joint set of poles enhances storage efficiency and reduces evaluation cost even further.

We will proceed by introducing a sample-based method to obtain a joint pole-expansion of all entries of the T-matrix:
\begin{equation}
\label{eq-pole-exp-T}
\mathbf{T}(k) \approx \mathbf{\hat T}(k) = \sum_n \frac{\boldsymbol{\mathcal{R}}_n}{k-k_n}  + \mathbf{BG}(k)\,,
\end{equation}
where $\mathbf{\hat T}(k)$ is matrix-valued due to the matrix-valued residues $\boldsymbol{\mathcal{R}}_n$ (and background $\mathbf{BG}(k)$).

To accommodate for the multimodal structure of $\mathbf{T}(k)$, the AAA-algorithm is modified for multidimensional transfer functions. In the machine learning community it is common to speak about tensor-valued quantities, which is why we denote this variant of the AAA-algorithm as \texttt{tensorAAA}. We start by using a generalized barycentric rational form \citep{hochman_fastaaa_2017, lietaert_automatic_2022}:
\begin{equation}
\label{eq:mat_baryrat}
\mathbf{R}(k) =  \frac{\sum_j \frac{w_j \mathbf{f}_j}{k-k_j}}{\sum_j \frac{w_j}{k-k_j}}\, .
\end{equation}
Notice how the weights $w_j$ remain scalar, such that all entries of $\mathbf{R}(k)$ share the same poles. This is a desired property, as it reflects the physical reality. The poles in the coefficients of $\mathbf{T}$ are manifestations of the scatterer's resonant states. As such, the contributions of a single resonant state across different multipoles should be assigned to a single pole. Indeed, the matrix-valued residues directly contain the \textit{normalized} radiating fields of the corresponding resonant states, as we show in Supplement III (`Normalized Resonant States from the Pole-Expansion of the T-matrix'). The only remaining modification is to adapt the weight optimization in the AAA-algorithm to minimize the error across all matrix entries. This modification is achieved by stacking the corresponding rows in the formulation of the optimization problem. The implementation is made publicly available with this publication as an update to our open-source Python package \texttt{diffaaable} \citep{Fischbach2026}.

\begin{figure*}[!htbp]
\centering
\includegraphics[width=0.8\linewidth]{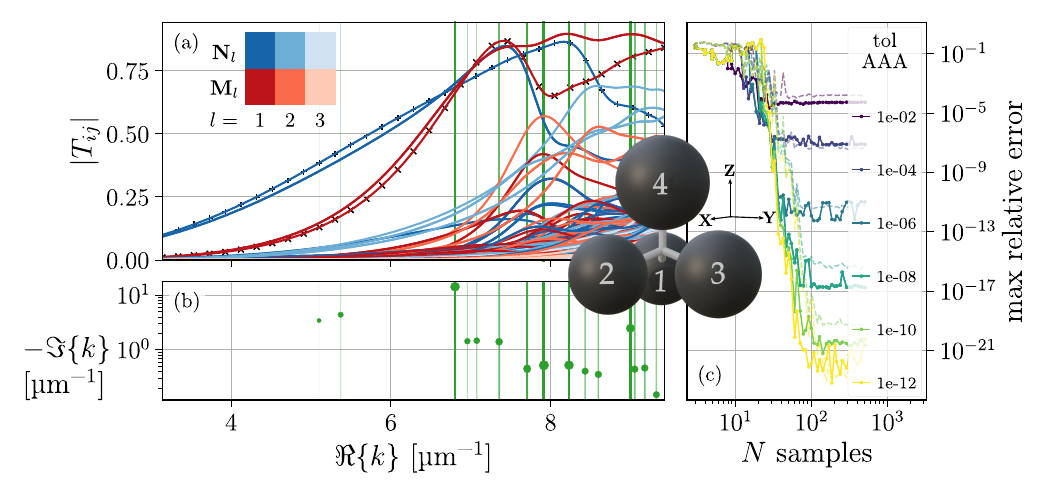}
\caption[]{Convergence study for the T-matrix of a scatterer consisting of four spheres with different radii (100, 110, 120, and \qty{130}{\nano\meter}) arranged in the order indicated in the inset (the white rods only serve as a guide to the eye). The spheres are placed at the corners of a regular tetrahedron with \qty{300}{\nano\meter} sidelength. As all geometrical symmetries are broken, the T-matrix is densely populated. (a) The magnitude of the individual entries of a converged pole-expansion (tol AAA = 10\textsuperscript{-8} and sufficient samples) of the T-matrix ($\ell_\mathrm{max} = 3$). Vertical green lines indicate the positions of the poles found by the AAA-algorithm, as shown in (b). The matrix elements are color-coded according to the multipolar type (electric: blues; magnetic: reds) and degree of the scattered field (see inset legend). In an attempt to reduce visual clutter, differenent incident multipoles are not discriminated, i.e., only the index $i$ in $T_ 
{ij}$ determines the color. To highlight the electric and magnetic z-oriented dipole coefficients, reference samples of the directly evaluated scattering coefficients are indicated as black pluses and crosses, respectively, exemplifying the excellent agreement. (b) Joint poles of all entries of the T-matrix (\textit{i.e.}, poles of the T-matrix) as found by the converged \texttt{tensorAAA} algorithm (green dots). The dot size vizualises the poles' influence on the real axis. It is computed as $\mathrm{area}_\mathrm{dot} \sim |\boldsymbol{\mathcal{R}}_n|_\mathrm{HS} / \Im\{k_n\}^2$. (c) Convergence of the pole-expansion with increasing number of samples $N$ used in the AAA-algorithm. The relative error is determined from the squared Hilbert-Schmidt norm of the difference between the pole-expansion and additional samples not seen by the AAA-algorithm according to Eq. (\ref{eq:error}). Different colors indicate different tolerance criteria for terminating the AAA iteration. Solid lines correspond to pole-expansions constructed from the \texttt{tensorAAA} algorithm, while dashed lines show results from applying the standard AAA-algorithm to each element of the T-matrix separately [this approach is further discussed in Supplement V (`Approximating Individual Matrix Elements Separately')].}
\label{fig-tetra}
\end{figure*}

In the following, we will treat two exemplary scatterers. First, we consider a synthetic object composed of four spheres arranged so that the object has no remaining geometrical symmetries (geometry shown in Figure~\ref{fig-tetra}). Specifically, the object consists of four differently sized spheres arranged on the corners of a tetrahedron. The chosen example is typical for benchmarking computational methods to obtain T-matrices \citep{Asadova_2025}, as quasi-analytic reference solutions are available and broken symmetries reveal potential flaws that might otherwise go unnoticed. A direct consequence of the lack of symmetries is the large number of populated T-matrix entries as shown in Figure~\ref{fig-tetra}(a). In the main text, we will treat the tetrahedron of spheres by considering it as a cluster of scatterers for which the isolated T-matrices are known analytically. Eq. (\ref{eq:cluster_tmat}) then allows us to compute the T-matrix of the full tetrahedron, including all multiple-scattering interactions. In Supplement IV (`FEM Convergence Study'), we further provide a convergence study of the pole-expansion when treating the tetrahedron directly with 3D FEM-simulations to obtain the T-matrix samples. All FEM-simulations in this work are performed with the commercially available toolbox JCMsuite \citep{JCMwave}.

In a second example, using 3D, axisymmetric FEM-simulations, we study the T-matrix of a dielectric cylinder (\qty{220}{\nano\meter} height $\times$ \qty{55}{\nano\meter} radius) as is commonly found in top-down fabricated metasurfaces (see Figure~\ref{fig-cylinder}) \citep{zhouEfficientSiliconMetasurfaces2017}. As a consequence of the cylindrical symmetry, the coupling between basis states of different $z$-projected total angular momentum (degree $m$) is suppressed \citep{Gladyshev_2020}, leading to a sparsely filled T-matrix shown in Figure~\ref{fig-cylinder}(a). The dipolar $m=0$ terms on the diagonal of the T-matrix, which correspond to the z-oriented electric and magnetic dipole polarizabilities \citep{zerulla_t-matrix_2023}, are indicated by pluses and crosses, respectively. We later use the fact that these almost coalesce at $k \approx$ \qty{12.93}{\per\micro\meter} to create an almost dual pair of quasi-bound states in the continuum (qBICs) by arranging the cylinders on a square grid with a carefully chosen lattice constant. Lastly, we will illustrate how the pole-expansion of T-matrices can serve as a valuable conceptual tool to shed light on the multipolar composition of these dual qBICs, which would otherwise remain hidden.

To simplify the discussion, we will limit the examples to isotropic materials without dispersion (\textit{i.e.}, $\varepsilon_r =$ 9 for the tetrahedron and $\varepsilon_r =$ 12+0.003$\ii$ for the cylinder). In systems with dispersion described by pole-based permittivity models, accumulation points arise. These, in principle, do not prohibit a pole-expansion. However, their treatment requires special attention and is discussed elsewhere \citep{broer_natural_2009, agsurmof}.

\subsection{Case Studies}

Figure~\ref{fig-tetra}(a) shows a converged pole-expansion of the T-matrix of the tetrahedron of spheres, while the green dots in Figure~\ref{fig-tetra}(b) [and the attached vertical lines in Figure~\ref{fig-tetra}(a) and (b)] mark the corresponding poles in the complex plane, which will be discussed in more detail later. To exemplify the excellent agreement of the pole-expansion with direct evaluations of the T-matrix, reference samples are indicated for the z-oriented electric and magnetic dipole as black pluses and crosses, respectively. The shown pole-expansion was generated with a tolerance setting of 10\textsuperscript{-8} and using $\sim 500$ samples. However, we demonstrate that significantly fewer samples are sufficient by investigating the convergence behavior of the pole-expansion.
We compare expansions constructed from increasing numbers of samples to reference solutions at frequencies not seen by the AAA-algorithm. Figure~\ref{fig-tetra}(c) shows how the quality of the pole-expansion depends on the number of frequency samples $N$ available to the AAA-algorithm.
The remaining relative error is determined from the squared Hilbert-Schmidt norm $|\mathbf{A}|^2_\mathrm{HS} = \mathrm{Tr}(\mathbf{A}^\dagger \mathbf{A})$ of the difference between the pole-expansion $\mathbf{\hat T}(k)$ and the actual $\mathbf{T}(k)$. $|\mathbf{T}|^2_\mathrm{HS}$ is proportional to the angle-averaged scattering cross section. Consequently, $|\mathbf{\hat T} -\mathbf{T}|^2_\mathrm{HS}$ is proportional to the angle-averaged spurious, \textit{i.e.}, erroneous, scattering cross section. The maximum relative error within the frequency interval of interest is defined as \citep{fernandez-corbaton_observation-based_2018}:
\begin{equation}
\label{eq:error}
\text{Max Relative Error} = \mathrm{max}_{k} \frac{1}{2}  \frac{\left|\mathbf{\hat T}(k)-\mathbf{T}(k)\right|_\mathrm{HS}^2}{\bigl|\mathbf{\hat T}(k)\bigr|_\mathrm{HS}^2 + \bigl|\mathbf{T}(k)\bigr|_\mathrm{HS}^2}\, .
\end{equation}
Here, the factor $1/2$ ensures that the relative error is bounded between 0 and 1.

\begin{figure*}[!htbp]
\centering
\includegraphics[width=0.8\linewidth]{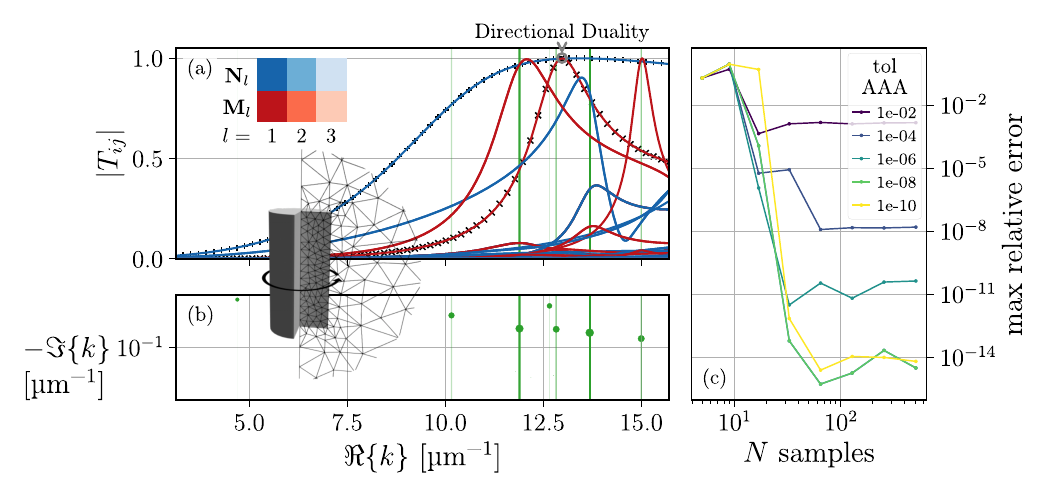}
\caption[]{Convergence study for a cylinder of \qty{220}{\nano\meter} height and \qty{55}{\nano\meter} radius, as depicted in the inset. (a) The spectra of all elements of the $\ell_\mathrm{max} = 3$ T-matrix are shown. As in Figure~\ref{fig-tetra} the z-oriented electric and magnetic dipole coefficients are indicated by black pluses and crosses, respectively. (b) The joint poles obtained from the \texttt{tensorAAA} are shown in analogy to Figure~\ref{fig-tetra}(b). (c) The convergence of the pole-expansion of the T-matrix with the number of considered frequency samples. We use the same error measure as in Figure~\ref{fig-tetra}.}
\label{fig-cylinder}
\end{figure*}

$4097=2^{12}+1$ equidistant reference samples were computed using \texttt{treams} \citep{beutelTreamsTmatrixbasedScattering2024} to perform the convergence study shown in Figure~\ref{fig-tetra}(c). From these reference samples, subsets to be passed to the AAA-algorithm are selected, each with a fixed stride. It is ensured that the first and last sample are always added to the subset to avoid a changing interval on which the pole-expansion is constructed. A maximum of 500 samples is selected, to make sure that sufficiently many reference samples remain, that have not been seen by the AAA-algorithm. From the selected samples the pole-expansion is constructed and the maximum relative error is evaluated according to Eq. (\ref{eq:error}). The pole-expansion quickly converges to an accurate representation of the frequency-dependent T-matrix. As indicated by the different colors, the final accuracy of the approximation depends, as expected, on the tolerance setting of the AAA-algorithm. The quadratic nature of the error measure explains the factor-of-two difference in the order of magnitude between the AAA tolerance and the final relative error.

In the investigated examples, the required number of samples to reach an approximation that introduces negligible error depends on the AAA tolerance (a stricter tolerance leads to slightly slower convergence) and on the error threshold limited by the numerical discretization error of the chosen simulation modality [see Supplement IV (`FEM Convergence Study')]. However, in Figure~\ref{fig-tetra}(c) we find that 50 samples are sufficient to push the error below the accuracy of the underlying scattering simulations. In Supplement VI (`Fluctuations in the Convergence of the \texttt{tensorAAA} Pole-Expansion') we investigate the fluctuations visible within the converged tails of the error curves in more detail and attribute the discrete jumps, particularly visible for $\text{tol}=10^{ -6}$, to the discrete fluctuation of the approximation degree. Comparing the \texttt{tensorAAA} algorithm (solid lines) with the standard AAA-algorithm applied to each element of the T-matrix separately (dashed lines) shows that both yield similar convergence behavior. Indeed, in our specific example, applying the AAA-algorithm to the individual matrix elements converges slower than the \texttt{tensorAAA}, but we did neither observe superior nor inferior convergence across all numerical experiments.

Furthermore, as shown in Figure~\ref{fig-tetra}(b), the \texttt{tensorAAA} algorithm by construction finds joint poles for all matrix elements (green dots). We show in Supplement V (`Approximating Individual Matrix Elements Separately'), that applying the standard AAA-algorithm to each element separately yields slightly shifted poles per element, accumulating in the vicinity of the \texttt{tensorAAA} joint poles. The errors when treating individual matrix elements are particularly pronounced for resonances that are almost decoupled from the respective matrix element. This observation highlights the advantage of the \texttt{tensorAAA} algorithm in providing a physically meaningful and data-efficient representation of the T-matrix.

Errors in the estimation of the resonance frequencies and residues of the dominant resonant state result in large errors in the rational approximation. Therefore, the \texttt{tensorAAA} intrinsically fits the dominant contributions most accurately. These are generally reflected by poles relatively close to the real frequency axis with large residues. To emphasize these particularly relevant poles, the dot size indicates the poles' influence on the real axis. It is computed as $\mathrm{area}_\mathrm{dot} \propto |\boldsymbol{\mathcal{R}}_n|_\mathrm{HS} / \Im\{k_n\}^2$. The corresponding resonances can be identified in the matrix elements $T_{ij}$. However, the circumstance that the T-matrix is densely populated obscures the correspondence between poles and resonance peaks. Due to its sparsity, the cylinder example provides a cleaner picture, presenting a clear mapping between poles and peaks.

Let us turn our attention to the second example of the dielectric cylinder. Due to its high degree of symmetry ($D_{\infty h}$), the T-matrix is sparse and contains recurring entries. This opportunity can be used to reduce the effort imposed by the convex minimization problem of optimizing the weights in the AAA iteration. In particular, we restrict ourselves to fitting only a single copy of any recurring element and then reconstructing the dense barycentric rational form from these. Strictly speaking, this procedure does not yield exactly the same fit as fitting the whole matrix, because recurring entries would, in the latter case, be weighted more strongly during the weight optimization. Nonetheless, we find that treating a single copy yields good results while keeping complexity low. In principle, this shifted weighting could be compensated for by rescaling the matrix elements according to their multiplicity before applying the AAA-algorithm.

Similar to the tetrahedron of spheres, the pole-expansion of the cylinder's T-matrix quickly converges as shown in Figure~\ref{fig-cylinder}(c). In this case, we compare to a finely resolved set of 513 reference solutions obtained from the same simulation modality. Consequently, we disregard the discretization error inherent to the simulation modality and purely focus on the convergence of the AAA approximation [in contrast to Supplement IV (`FEM Convergence Study')]. As with the tetrahedron before, we observe rapid convergence until a remaining error, dictated by the AAA tolerance, is reached.
Due to the symmetry-enforced sparsity of the T-matrix, the correspondences between poles and resonance peaks in the matrix elements $T_{ij}$ are more obvious, as shown in Figure~\ref{fig-cylinder}(a) and (b). We would like to mention that the resonance peaks in the matrix elements $T_{ij}$ do not align exactly with the real part of the poles. This is caused by Fano-like interference effects from neighboring resonances and the background term.

\begin{figure*}[!htbp]
\centering
\includegraphics[width=0.8\linewidth]{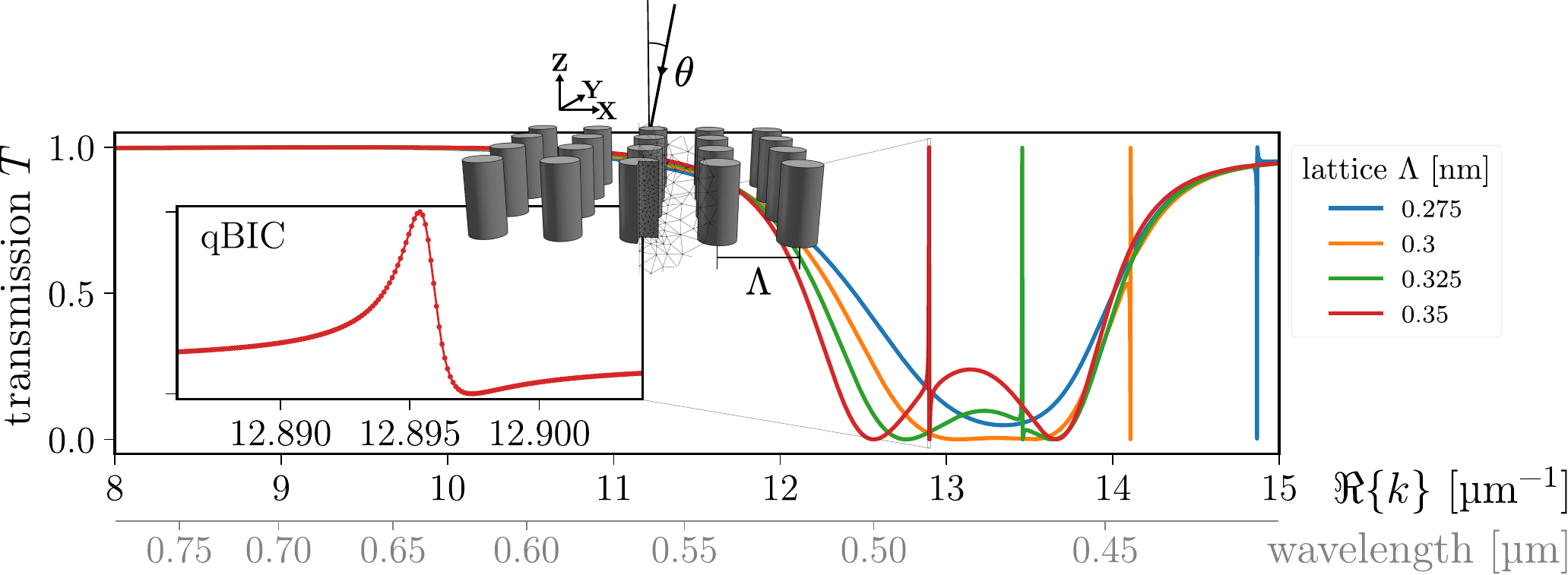}
\caption[]{Symmetry-protected quasi-bound states in the continuum (qBICs) of an infinite square lattice of cylinders with different lattice spacing $\Lambda$ under slightly off-axis TM illumination (component of the wavevector in the x-direction $k_{x} =$ \qty{0.006}{\per\micro\meter}; which corresponds to an incidence angle $\theta$ of 0.04297° to 0.02292° within the shown frequency window). The pole-expansion of the T-matrix enables a dense frequency sampling of the transmission spectrum $T$, which is needed to faithfully resolve the qBIC lineshape (see inset). The computational advantage is particularly pronounced in the presented example, as it enables leveraging the cylindrical symmetry of the meta-atom despite the lattice breaking the symmetry.}
\label{fig-moving-bics}
\end{figure*}

\subsection{Computational Efficiency}

Here, we will briefly discuss the computational cost advantage unlocked by the pole-expansion. Furthermore, we will demonstrate the reduced memory requirements of representing the frequency dependence using shared poles and matrix-valued residues rather than a finely resolved set of discrete frequency samples.
The computational effort associated with the AAA pole-expansion is twofold:

\begin{enumerate}
\item Finding the approximation in barycentric rational form.
\item Evaluating the barycentric rational approximant at the desired frequencies.
\end{enumerate}

For scalar-valued transfer functions, step 1 has negligible cost. When expanding matrix valued transfer functions with upwards of 900 coefficients (a T-matrix with $\ell_\mathrm{max}=$ 3 has 30 $\times$ 30 entries) the singular value decomposition used to find the weights in the AAA iteration becomes a bottleneck, resulting in an overall complexity of $O(b^2Nn^3)$, where $b$ is the size of the vector spherical wave basis, N is the number of frequency samples fed to the \texttt{tensorAAA} and $n$ is the number of required poles.

Several methods have been proposed to improve the scaling of the matrix-valued AAA-algorithm \citep{hochman_fastaaa_2017}. Implementing the iteratively updated QR-decompositions as suggested in \citep{lietaert_automatic_2022}, the maximum AAA evaluation time reduces from approximately \qty{5}{\min} (tetrahedron, $\ell_\mathrm{max}=$ 3 with 105 frequency samples) to less than a second.

After the pole-expansion has been obtained, the costs per frequency sample are negligible at approximately \qty{0.106}{\milli\second} and \qty{0.147}{\milli\second} \footnote{All computations were performed on a 2xAMD EPYC 7453 compute node.} for the cylinder and tetrahedron example, respectively. Single full-wave evaluations for both examples take multiple tens of seconds. Thus, the pole-expansion is vastly more efficient in cases that require very fine frequency sampling, for example a lattice of cylinders sustaining a quasi-bound state in the continuum (qBIC), as shown in Figure~\ref{fig-moving-bics}. Resolving the narrow qBICs, as shown by the red dots in the inset, with equidistant frequency sampling required 90 thousand evaluations in the shown frequency window from \qty{8}{\per\micro\meter} to \qty{15}{\per\micro\meter}, making such studies very costly if the T-matrix had to be evaluated explicitly using a full-wave solver at every frequency. In contrast, only approximately 40 FEM evaluations are required when using the pole-expansion [as demonstrated in Figure~\ref{fig-cylinder}(c)].

Similarly, the pole representation requires storing $2(b^2+1)n$ floating-point numbers (the factor of 2 results from the real and imaginary parts), while discrete frequency sampling requires $2b^2N_\text{dense}$ floating-point numbers, where $N_\text{dense}$ is the number of samples needed to reach the desired spectral density. Thus, the ratio of frequency samples to resonances directly indicates data efficiency gains:
\begin{equation}
\text{Data Efficiency Gain} = \frac{2b^2N_\text{dense}}{2(b^2+1)n} \stackrel{b^2 \gg 1}{\approx} \frac{N_\text{dense}}{n}\, .
\end{equation}
We provide a utility package \texttt{baryTmat} with this publication, which enables storage and retrieval in a format compatible with \citep{Asadova_2025}. It directly saves the barycentric rational form as presented in Eq. (\ref{eq:mat_baryrat}). The resulting file thus contains only the directly obtained sparse frequency samples, augmented with the weights $w_j$.

\subsection{Revealing Multipolar Contents of Quasi-Dual BICs}

Lastly, we wish to highlight that, in addition to its computational advantages, our method provides valuable physical insights into resonant scattering phenomena. In metasurfaces with a square lattice, there exist at least two types of BICs: a symmetry-protected BIC at normal incidence and an accidental BIC at off-angle illumination. In this work, we focus exclusively on the first type. However, since a true BIC does not couple to incident light and therefore cannot be excited, we have so far considered high-Q resonances in at small angles of incidence. However, we will see in this section, that our method allows us to directly investigate the true BICs. To demonstrate this, we are going to dig a little deeper into the cylinder metasurface introduced above, reproducing a phenomenon previously predicted by \cite{evlyukhin_polarization_2021}, where two symmetry-protected BICs (dominated by the z-oriented electric and magnetic dipoles respectively) are forced to coincide for an appropriately chosen lattice constant. We will inspect the multipolar origin of these BICs by leveraging the pole-expansion of the lattice T-matrix, which will allow us to explore their characteristic features and observe their almost dual correspondence.

In Figure~\ref{fig-moving-bics} we have observed the spectral position of the electric dipole dominated qBIC of the lattice of cylinders red-shifting to lower energies with increasing lattice constant $\Lambda$. And so does the corresponding BIC. Reference \cite{evlyukhin_polarization_2021} explains this shift for dipolar meta-atoms from the coincidence of the dipolar lattice sum and the inverse of the $z$-oriented electric/magnetic polarizabilities ($\alpha_{zz}^\mathrm{ee}$ and $\alpha_{zz}^\mathrm{mm}$) \citep{evlyukhin_polarization_2021}. Going back to Figure~\ref{fig-cylinder}(a), we recall that the cylinder considered here features a point at which the $z$-oriented magnetic and electric diagonal entries of the dipolar T-matrix ($T_{zz}^\mathrm{ee} = T_{\ell=\ell^\prime=1,m=m^\prime=0,p=p^\prime=1}$, $T_{zz}^\mathrm{mm} = T_{\ell=\ell^\prime=1,m=m^\prime=0,p=p^\prime=0}$) are almost equal [marked with a gray circle; henceforth we will call this directional duality, as equal electric and magnetic scattering coefficients mark the condition for duality (see Ref. \citep{fernandez-corbatonDualChiralObjects2015}; further, see Eq. (3) of Ref. \citep{rahimzadeganCoreShellParticlesBuilding2018}: achiral scatterers $\mathbf{T}^\mathrm{em} = \mathbf{T}^\mathrm{me} = 0$ are dual-symmetric if $\mathbf{T}^\mathrm{ee} = \mathbf{T}^\mathrm{mm}$). However, this condition is only fulfilled for the z-oriented dipoles, but not for the in-plane dipoles -- hence the added \textit{directional}]. From the correspondence between the dipolar T-matrix and the dipole polarizability tensors \citep{zerulla_multi-scale_2022}, we find:
\begin{equation}
\alpha_{zz}^\mathrm{ee} = \frac{\ii k_0^3}{6\pi} T_{zz}^\mathrm{ee} \quad \text{and} \quad \alpha_{zz}^\mathrm{mm} = \frac{\ii k_0^3}{6\pi} T_{zz}^\mathrm{mm}
\end{equation}
From that, it follows that, if the scatterers were purely dipolar, the directional duality would allow forcing two BICs to coincide at the point of directional duality. Indeed, we demonstrate in Supplement VII (`Dipolar Condition for Coincident BICs'), that neglecting higher multipoles, coincident dipolar electric and magnetic BICs exist for a lattice constant of $\Lambda_\mathrm{dipole}\approx$  \qty{353.8}{\nano\meter}.

As introduced in Section~\ref{sec:tmat}, the lattice T-matrix ($\mathbf{T}_\mathbf{k_\parallel}(k)$; defined by Eq. (17) of Ref. \citep{beutel_efficient_2021}) captures all interactions of a single meta-atom with its surrounding infinite lattice. As the BICs of interest arise at normal incidence, $\mathbf{T}_{\mathbf{k_\parallel}=0}(k)$ contains the BICs, being collective effects of the lattice. At first sight, this might seem contradictory, because BICs do not couple to the far-field by definition and T-matrices relate incident and scattered VSWs (\textit{i.e.}, far-field amplitudes). However, the periodic repetition of scattered VSWs can collectively form evanescent waves that decouple from the radiation continuum through destructive interference outside the open diffraction channels. Therefore, also resonances that are fully decoupled from the radiation continuum (\textit{i.e.}, BICs) lead to singularities in the lattice T-matrix. The mechanism can be understood similarly to the excitation of BICs from the near field \citep{corrielli_observation_2013, hsu_bound_2016, abujetas_near-field_2021, van_hoof_unveiling_2021}.

\begin{figure}[!htbp]
\centering
\includegraphics[width=1\linewidth]{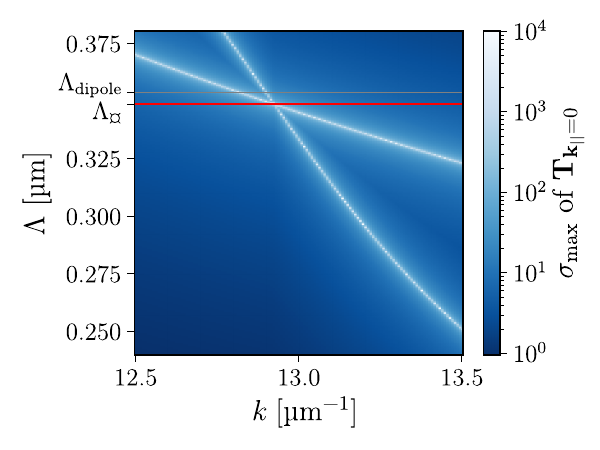}
\caption[]{Finetuning of the lattice constant to enforce the coincidence of BICs, considering the effect of higher multipoles. When supporting a BIC lattice-resonance, the lattice T-matrix $\mathbf{T}_\mathbf{k_\parallel}(k)$ becomes singular at a real frequency. As such, we can track the BICs through the largest singular value $\sigma_{\text{max}}$ of $\mathbf{T}_\mathbf{k_\parallel}(k)$. The gray horizontal line indicates the lattice constant $\Lambda_\mathrm{dipole}$ predicted from dipolar considerations only, while $\Lambda_\gem \approx$ \qty{348.64}{\nano\meter} marks the lattice constant for which the two quasi BICs actually coincide.}
\label{fig-finetuning}
\end{figure}

Consistent with observations in Ref. \citep{evlyukhin_polarization_2021}, the non-negligible contributions from higher multipoles shift the quasi-dual BICs slightly away from the frequency where the cylinder exhibits directional duality. Consequently, the lattice constant $\Lambda$ for which the BICs coincide differs slightly from $\Lambda_\mathrm{dipole}$. As shown in Figure~\ref{fig-finetuning}, the representation of $\T(k)$ in terms of its poles allows us to efficiently eliminate the remaining frequency mismatch numerically, by fine-tuning the lattice constant to $\Lambda_\gem \approx$ \qty{348.64}{\nano\meter}. To track the BICs, we consider the largest singular value $\sigma_{\text{max}}$ of $\mathbf{T}_{\mathbf{k_\parallel}}(k)$ at normal incidence, which diverges whenever $\mathbf{T}_{\mathbf{k_\parallel}}(k)$ becomes singular.

\begin{figure}[!htbp]
\centering
\includegraphics[width=1\linewidth]{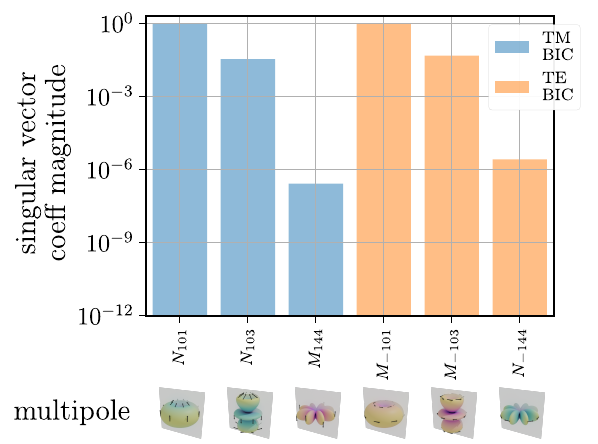}
\caption[]{Multipolar contents of the quasi-dual BICs at normal incidence.
We analyze the contents of the left singular vectors of the residues of the two BICs shown in Figure~\ref{fig-finetuning} at the point where they coincide. Only multipoles that do not radiate along the $z$-direction contribute to the BICs, which is reflected in the vanishing coefficients of all multipoles with $m=1$. The insets illustrate the far-field radiation patterns of the multipoles constituting the two BICs, which are non-radiating along the $z$-direction. The E-field magnitude is represented as the distance of the surface from the origin. The orientation of the E-field is indicated by the black arrows attached to the surface. The $zx$-plane is indicated in gray to help visualize the E-field parity upon y-reflection, which determines the first multipole index in the notation of \cite{sadrieva_multipolar_2019}. The T-matrices obtained from FEM simulations were symmetrized with respect to reflections at the $xy$-plane. Otherwise, additional multipoles appear in the BICs, as is shown in Supplement VIII (`Restoring xy-Plane Mirror Symmetry'). We attribute the lack of reflection symmetry in the T-matrix to the numerical discretization (\textit{i.e.}, the meshing) not respecting said symmetry.}
\label{fig-multipoles}
\end{figure}

For isolated objects, the singular vectors of the residues of $\T(k)$ contain the radiated fields of the resonant state expressed in the basis of outgoing VSWs [as discussed in Supplement III (`Normalized Resonant States from the Pole-Expansion of the T-matrix')]. To investigate the multipolar contents in the current example of coinciding BICs, we therefore consider the corresponding residues in the pole-expansion of the lattice T-matrix. In Figure~\ref{fig-multipoles} we plot the singular vectors (there is only one significant singular value per residue), revealing an almost dual correspondence between the two coinciding BICs. In other words, the multipolar composition of one of the BICs is almost identical to the other, but with electric and magnetic multipoles interchanged. This justifies the name \textit{quasi-dual BICs}. The correspondence is not perfect, as the chosen meta-atom is not actually dual, but rather exhibits large directional duality. The resulting difference is most apparent in the strength of the hexadecapoles $M_{144}$ and $N_{ -144}$ (in the notation of Ref. \citep{sadrieva_multipolar_2019}).

In contrast to the rest of this manuscript, which uses Jackson's complex-valued vector spherical waves \citep{jackson_classical_1999}, we present in this figure the multipolar composition in terms of real multipoles, as used by Bohren and Huffman \citep{bohren_absorption_1998}, but using the indices as in Ref. \citep{sadrieva_multipolar_2019}. This enables the direct comparison of our results to the group theoretical predictions of \cite{sadrieva_multipolar_2019}, which have shown that in a lattice with $D_{4h}$ symmetry, $\mathrm{TE}_1$ (\textit{i.e.}, magnetic dipole-like) BICs at normal incidence ($\Gamma$-point) are composed of multipoles, that transform according to the irreducible representation (irrep) $A_{2g}$. At the same time, $\mathrm{TM}_4$ (\textit{i.e.}, electric dipole-like) BICs consist of multipoles transforming according to the irrep $A_{2u}$. Our findings are in perfect agreement with these predictions (see Tables I and II of Ref. \citep{sadrieva_multipolar_2019} for an overview of which multipoles should contribute to the BICs). As such, the BICs do not contain multipoles with $m=1$, which are the only multipoles that radiate along the z-axis -- the only open diffraction channel \citep{sadrieva_multipolar_2019}.

The transmissivity of the lattice with coinciding qBICs under very mildly off-angle ($k_x=\,$\qty{1e-4}{\per\micro\meter}) plane wave illumination is shown in Supplement IX (`Off-axis Illumination'). It reveals that the lack of duality for in-plane-oriented dipoles leads to drastically different linewidths for the two qBICs. This configuration results in interesting transmission spectra for circularly polarized plane waves, with multiple nested Fano-like features. The lattice resonances of the isolated meta-atom form a slowly varying background, on top of which the broader TM qBIC creates a Fano-like feature that itself hosts the narrower TE qBIC as a nested Fano resonance.

\section{Conclusion}

In the present work, we demonstrate a method based on a matrix-valued variant of the AAA-algorithm (made available as \texttt{tensorAAA} in \texttt{diffaaable}) to obtain the pole-expansion of the T-matrix with only a small number of frequency evaluations. We introduce two examples to highlight several aspects of this alternative frequency representation, including its excellent convergence, its benefits in evaluation time, and reduced storage requirements.
We concluded our discussion by highlighting the occurrence of quasi-dual BICs and their multipolar composition as an example of an interesting physical phenomenon particularly well described by the pole-expansion of the (lattice) T-matrix. Our analysis sheds light on the multipolar structure of the BICs, showing an almost dual correspondence between the co-localized resonances. Our findings are, in principle, applicable to any wave phenomena that can be described by transition-matrix formalisms, ranging from acoustic scattering and seismic waves to scattering in quantum systems.
In summary, the pole-expansion of the T-matrix is an exciting tool for nanophotonic scattering and beyond, offering numerical advantages while providing valuable physical insight.
\begin{acknowledgments}
J.D.F., L.R., and P.G. acknowledge support from the Karlsruhe School of Optics and Photonics (KSOP).
J.D.F. and C.R. acknowledge financial support by the Helmholtz Association in the framework of the innovation platform ``Solar TAP''.
L.R., P.G., and C.R. acknowledge financial support by the Deutsche Forschungsgemeinschaft (DFG, German Research Foundation) -- Project-ID 258734477 -- SFB 1173.
P.G. and C.R. are part of the Max Planck School of Photonics, supported by the Bundesministerium für Forschung, Technologie und Raumfahrt, the Max Planck Society, and the Fraunhofer Society.
K.F. gratefully acknowledges support from the Alexander von Humboldt Foundation.
F.B., F.B. and S.B. acknowledge funding by the Deutsche Forschungsgemeinschaft under Germany's Excellence Strategy -- The Berlin Mathematics Research Center MATH+ (EXC-2046/1, Project-ID 390685689) and by the Bundesministerium für Forschung, Technologie und Raumfahrt (BMFTR Forschungscampus MODAL, Project-ID 05M20ZBM).

The datasets generated during and/or analyzed during the current study are available in the GitHub repository, \href{http://www.github.com/tfp-photonics/pole-expansion-t-matrix}{www.github.com/tfp-photonics/pole-expansion-t-matrix}. Further, the T-matrices generated during the current study are available in the T-matrix database accompanying Ref. \citep{Asadova_2025}.
\end{acknowledgments}
\bibliography{main.bib}

\balancecolsandclearpage
    
\setcounter{section}{0}
\setcounter{subsection}{0}
\setcounter{figure}{0}
\setcounter{table}{0}
\setcounter{equation}{0}

\renewcommand{\thesection}{\Roman{section}}
\renewcommand{\thesubsection}{S\arabic{section}.\arabic{subsection}}
\renewcommand{\thefigure}{S\arabic{figure}}
\renewcommand{\thetable}{S\arabic{table}}
\renewcommand{\theequation}{S\arabic{equation}}

\section*{Supplementary Information}

\renewcommand{\pol}{s}
\section{Sphere with Chiral Shell}

In the main text we have introduced Maxwell's equations only considering achiral media. In Figure 1 of the main text a chiral core-shell particle provides a convenient example of a T-matrix with a small number of Mie-resonances (in the considered frequency interval), that appear separately in the electric-electric and magnetic-magnetic dipole terms, while all resonances appear in the electric-magnetic coupling terms. For the spherically symmetric geometry, the electric magnetic coupling terms only appear due to a nonzero pasteur parameter $\kappa$ in the more general isotropic macroscopic Maxwell equations that account for chiral media:

\begin{equation}
\nabla\times
\begin{pmatrix}
\mathbf{E}(k, \rr) \\
Z_0\mathbf{H}(k, \rr)
\end{pmatrix}=k
\begin{pmatrix}
\kappa & \mathrm{i}\mu(k) \\
-\mathrm{i}\varepsilon(k) & \kappa
\end{pmatrix}
\begin{pmatrix}
\mathbf{E}(k, \rr) \\
Z_0\mathbf{H}(k, \rr)
\end{pmatrix}.
\end{equation}

The considered core-shell particle consists of an achiral core ($r_\mathrm{core} = \qty{0.5}{\micro \meter}$; $\varepsilon = 12 \varepsilon_0$; $\mu = \mu_0$; $\kappa = 0$) with a shell ($r_\mathrm{shell} = \qty{0.6}{\micro \meter}$; $\varepsilon = 12 \varepsilon_0$; $\mu = \mu_0$; $\kappa = 2\times10^5$) made of a fictitious chiral material. Its T-matrix is sampled and expanded into poles. It should only be considered as an illustrative example to introduce the sample-based approach to the pole-expansion of the T-matrix. Due to the spherical symmetry, the dipolar T-matrix is fully described by the electric, magnetic and cross-coupling coefficients, that are shown in the figure.

\section{Analytic Continuation of the VSW Decomposition Integrals}

For the T-matrix elements $T_{ij}$ to be efficiently expressed as a sum over pole contributions, an analytic (\textit{i.e.}, holomorphic) continuation of $T_{ij}$ in $k$ should exist. Considering the decomposition integral [Eq. (9) of the main text], this is not immediately obvious, due to the complex conjugation, which is not holomorphic \citep{BetzQuadratic_2023}. To find an appropriate continuation, we need to identify an expression that is equivalent on the real axis, but does not contain the conjugation of any $k$-dependent expression. For this purpose we will consider the explixit expressions for the VSWs (in the conventions of \texttt{treams} \citep{beutelTreamsTmatrixbasedScattering2024}):

\begin{equation}
\begin{aligned}
\Mlmn &=\genbessel\boldsymbol{X}_{\ell m}(\theta,\varphi),
\\
\Nlmn &=
\\
&\left(\genbesselprime +\frac{\genbessel}{kr}\right)&\boldsymbol{Y}_{\ell m}(\theta,\varphi)
\\
&+\sqrt{\ell(\ell+1)}\frac{\genbessel}{kr}&\boldsymbol{Z}_{\ell m}(\theta,\varphi).
\end{aligned}
\end{equation}
Here, $\boldsymbol{X}_{\ell m}(\theta,\varphi)$, $\boldsymbol{Y}_{\ell m}(\theta,\varphi)$, $\boldsymbol{Z}_{\ell m}(\theta,\varphi)$ are the vector spherical harmonics with order $\ell$ and degree $m$, $\rr$ is the location vector with components $(r,\theta,\varphi)^\mathsf{T}$ in spherical coordinates.

Instead of conjugating $\mathbf{\Psi}_{\ell ms}^{(n)}(k, \mathbf{r})$, we exchange the outgoing for incoming spherical Hankel functions and conjugate the remaining vector spherical harmonics, which have only angular dependence: 
\begin{equation}
\begin{aligned}
\tMlm{3} &=\genbesseli{4}\boldsymbol{X}_{\ell m}^{*}(\theta,\varphi)\, , 
\\
\tNlm{3} &=
\\
&\left(\genbesselprimei{4} +\frac{\genbesseli{4}}{kr}\right)&\boldsymbol{Y}_{\ell m}^{*}(\theta,\varphi)
\\
&+\sqrt{\ell(\ell+1)}\frac{\genbesseli{4}}{kr}&\boldsymbol{Z}_{\ell m}^{*}(\theta,\varphi)\, .
\end{aligned}
\end{equation}

Due to the relation $z_l^{(3)*}(kr) = \genbesseli{4}$ for real $kr$ (\textit{i.e.}, the complex conjugate relation between the Hankel functions of first and second kind), we get $\tNlm{3} = \cNlm{3}$ and $\tMlm{3} = \cMlm{3}$, respectively. Note, how $\tNlm{3}$ and $\tMlm{3}$ fulfill incoming boundary conditions as opposed to outgoing boundary conditions met by $\Nlm{3}$ and $\Mlm{3}$. Combining these expressions with appropriate solvers to evaluate the analytic continuation of $\mathbf{E}_{\mathrm{sca}, \ell^\prime m^\prime s^\prime}(k, \mathbf{r})$ enables the evaluation of the analytic continuation of the matrix elements $\tilde T_{\ell ms, \ell^\prime m^\prime s^\prime} (k)$. As a result of the decomposition, integrals are independent of the chosen integration surface (given that the integration surface contains all inhomogeneities). Moreover, due to the uniqueness of the analytic continuation, the found analytic continuation itself is independent of the integration surface. In this work, we use the commercially available finite element solver JCMsuite. The presented changes to the VSW decomposition integrals have been adopted in JCMsuite version 6.2.8 and later. \textit{I.e.}, up-to-date versions of JCMsuite use $\tNlm{3}$ and $\tMlm{3}$ instead of $\cMlm{3}$ and $\cMlm{3}$.

\section{Normalized Resonant States from the Pole-Expansion of the T-Matrix}

Here, we derive how normalized resonant states [also called quasinormal modes (QNMs)] can be obtained directly from the residues of the pole-expansion of the T-matrix.

\subsection{Pole Response Normalization}

In \citep{Bai_2013}, the authors introduce a QNM normalization \textbf{not} reliant on a volume integral, which is in contrast to other QNM normalization schemes. Nonetheless, the resulting normalization is known to be equivalent to the other approaches. It is discussed in the review paper \citep{Sauvan_2022} and formulated in terms of the response to an excitation by a dipole source. The normalization reads as \citep{Sauvan_2022}:
\begin{equation}
\label{eq-pole-response-normalization}
\tilde{\mathbf{E}}_m(\mathbf{r})=\lim_{\omega\to\tilde{\omega}_m}\sqrt{\frac{\tilde{\omega}_m-\omega}{\omega\mathbf{p}\cdot\mathbf{E}_\mathrm{sca}(\mathbf{r}_\mathrm{p},\omega, \mathbf{r}_\mathrm{p})}}\mathbf{E}_\mathrm{sca}(\mathbf{r},\omega, \mathbf{r}_\mathrm{p})\, ,
\end{equation}
Where $\tilde{\mathbf{E}}_m(\mathbf{r})$ is the normalized QNM-field, $\tilde{\omega}_m$ is the complex valued resonance frequency, $\mathbf{p}$ is the source dipole moment (with frequency $\omega$ and located at $\mathbf{r}_\mathrm{p}$), and $\mathbf{E}_\mathrm{sca}(\mathbf{r},\omega, \mathbf{r}_\mathrm{p})$ is the scattered field in response to the excitation.

\subsection{Pole-Expansion of the T-Matrix}

Here, we use the technique mentioned above to normalize modes found via the pole-expansion of the T-matrix. Note that we switch to the pole-expansion in terms of $\omega$ instead of $k$ for this derivation to increase consistency with the literature on QNMs:

\begin{equation}
\mathbf{T}(\omega) = \sum_m \frac{\res}{\omega-\tilde{\omega}_m}  + \mathbf{BG}(\omega)\, .
\end{equation}

As mentioned in the main text, for expansions in angular frequency $\omega=c_0 k$ the residues relate to those in wavenumber $k$ as $\res = \boldsymbol{\mathcal{R}}_n^{(k)} / c_0$.
Close to a pole (which is where we have to evaluate the normalization), the corresponding contribution is dominant:
\begin{equation}
\label{eq-pole-exp-T-close}
\lim_{\omega\to\tilde{\omega}_m} \left | \mathbf{T}(\omega) - \frac{\res}{\omega-\tilde{\omega}_m} \right |_\mathrm{HS} = 0 \, .
\end{equation}

\subsection{Pole Response Normalization based on the T-Matrix Residues}

\begin{figure}[!htbp]
\centering
\includegraphics[width=0.7\linewidth]{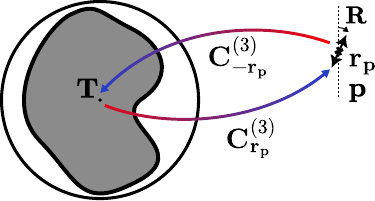}
\caption[]{Schematic overview of the pole response normailzation leveraging the T-matrix. A dipole source at an arbitrary position illuminates the scatterer. The incident field is calculated using rotation and translation-addition theorems of VSWs. The scattered field is then evaluated at the position of the dipole (once again using rotation and translation-addition theorems of VSWs) and used to normalize the QNM fields. In the derivation we show that the final expression for the normalized QNMs only depends on the resonance frequency and the corresponding residue of the T-matrix.}
\label{fig-backandforth}
\end{figure}

We will have to evaluate the scattered field resulting from a dipole moment $\mathbf{p}$ at $\mathbf{r}_\mathrm{p}$ for which we define the operator $\mathcal{E}_\mathbf{r}(\omega)$, which, applied to the corresponding vector of VSW coefficients, evaluates the field at position $\mathbf{r}$:
\begin{equation}
\mathbf{E}_\mathrm{sca}(\mathbf{r}, \omega, \mathbf{r}_\mathrm{p}) = \mathcal{E}_\mathbf{r}(\omega)\mathbf{q}(\omega, \mathbf{r}_\mathrm{p})\, .
\end{equation}
Here, $\mathbf{q}(\omega, \mathbf{r}_\mathrm{p})$ is the vector of VSW coefficients. The functional dependence on $\mathbf{r}_\mathrm{p}$ indicates that the coefficients depend on the position of the dipole source. It should \textit{not} be understood as `evaluating it at point $\mathbf{r}_\mathrm{p}$'.

Using this operator notation to evaluate the field at the position of the dipole via $\mathcal{E}_\mathbf{r_p}(\omega)$ we can formulate Eq. (\ref{eq-pole-response-normalization}) as:
\begin{equation}
\tilde{\mathbf{E}}_m(\mathbf{r})=\lim_{\omega\to\tilde{\omega}_m}\sqrt{\frac{\tilde{\omega}_m-\omega}{\omega\mathbf{p}\cdot \mathcal{E}_{\mathbf{r}_\mathrm{p}}(\omega)\mathbf{q}(\omega, \mathbf{r}_\mathrm{p})}}\mathcal{E}_\mathbf{r}(\omega)\mathbf{q}(\omega, \mathbf{r}_\mathrm{p})\, .
\end{equation}
And further using Eq. (\ref{eq-pole-exp-T-close})
\begin{equation}
\begin{aligned}
\tilde{\mathbf{E}}_m(\mathbf{r})&=\\\lim_{\omega\to\tilde{\omega}_m}&\sqrt{\frac{\tilde{\omega}_m-\omega}{\omega\mathbf{p}\cdot \mathcal{E}_{\mathbf{r}_\mathrm{p}}(\omega) \frac{\res}{\omega-\tilde{\omega}_m} \mathbf{a}(\omega, \mathbf{r}_\mathrm{p}) }}\mathcal{E}_\mathbf{r}(\omega)\frac{\res}{\omega-\tilde{\omega}_m} \mathbf{a}(\omega, \mathbf{r}_\mathrm{p}).
\end{aligned}
\end{equation}
The $\omega -\tilde{\omega}_m$ factors cancel leaving us with:
\begin{equation}
\label{eq-freq-independent}
\begin{aligned}
\tilde{\mathbf{E}}_m(\mathbf{r})&=\lim_{\omega\to\tilde{\omega}_m}\frac{\mathcal{E}_\mathbf{r}(\omega)\res \mathbf{a}(\omega, \mathbf{r}_\mathrm{p})}{\sqrt{- \omega\mathbf{p}\cdot \mathcal{E}_{\mathbf{r}_\mathrm{p}}(\omega) \res \mathbf{a}(\omega, \mathbf{r}_\mathrm{p}) }}\\ \\ &= \mathcal{E}_\mathbf{r}(\tilde{\omega}_m) \underbrace{\frac{\res \mathbf{a}(\tilde{\omega}_m, \mathbf{r}_\mathrm{p})}{\sqrt{- \tilde{\omega}_m\mathbf{p}\cdot \mathcal{E}_{\mathbf{r}_\mathrm{p}}(\tilde{\omega}_m) \res \mathbf{a}(\tilde{\omega}_m, \mathbf{r}_\mathrm{p}) }}}_{\mathbf{q}_{m}}\, .
\end{aligned}
\end{equation}
Due to the cancelation of the $\omega -\tilde{\omega}_m$ factors we were able to take the limit explicitly. Hence, we will drop the $\omega$ dependence in the following, as we consider it fixed to $\tilde{\omega}_m$. Here, we define $\mathbf{q}_m$ as the radiating normalized QNM fields expressed as coefficient vectors in terms of outgoing VSWs.
Now, let us express the incident field caused by an arbitrarily oriented and positioned dipole in VSWs:
\begin{equation}
\label{eq-dipole-incidence}
\mathbf{a}(\mathbf{r}_\mathrm{p}, \mathbf{p}) = \mathbf{C}_{-\mathbf{r}_\mathrm{p}}^{(3)} \mathbf{R} \mathbf{d} \frac{|\mathbf{p}|}{p_0}\, ,
\end{equation}
where $\mathbf{C}_{ -\mathbf{r}_\mathrm{p}}^{(3)}$ is the translation from singular VSWs centered at $\mathbf{r}_\mathrm{p}$ to regular VSWs centered at the origin. $\mathbf{R}$ is the rotation operator encapsulating the orientation of the dipole, and $\mathbf{d}$ is the vector of VSW coefficients of the z-oriented electric dipole (\textit{i.e.}, $\mathbf{d} = p \delta_{\ell,1}\delta_{m,0}\delta_{p,1}$). Further, $p_0 = \sqrt{6 \pi}\frac{\varepsilon_0}{k_0^3}$ is the dipole moment generating that VSW.

In addition, we use the fact that out of all regular VSWs only the electric z-oriented dipole has a non-vanishing z-oriented E-field at the origin. This lets us evaluate the E-field of an arbitrary scattered field in VSW coefficients $\mathbf{q}$ projected onto the dipole by isolating the coefficient of the z-oriented electric dipole after expanding the scattered fields into regular VSWs around the origin:
\begin{equation}
\label{eq-dipole-trick}
\mathbf{p}\cdot \mathcal{E}_{\mathbf{r}_\mathrm{p}} \mathbf{q} = |\mathbf{p}| \frac{\mathrm{i}}{\sqrt{6\pi}} \mathbf{d}^\mathsf{T} \mathbf{R}^{-1} \mathbf{C}_{\mathbf{r}_\mathrm{p}}^{(3)} \mathbf{q}\, .
\end{equation}
Plugging Eqs. (\ref{eq-dipole-incidence}) and (\ref{eq-dipole-trick}) into Eq. (\ref{eq-freq-independent}) we get:
\begin{equation}
\mathbf{q}_m(\mathbf{r}) = \frac{\res \mathbf{C}_{-\mathbf{r}_\mathrm{p}}^{(3)} \mathbf{R} \mathbf{d} }{ \sqrt{- \tilde{\omega}_m \frac{\mathrm{i} \varepsilon_0}{k_0^3} \mathbf{d}^\mathsf{T} \mathbf{R}^{-1} \mathbf{C}_{\mathbf{r}_\mathrm{p}}^{(3)} \res \mathbf{C}_{-\mathbf{r}_\mathrm{p}}^{(3)} \mathbf{R} \mathbf{d} }}\, .
\end{equation}

We proceed by expressing the residue $\res$ in terms of its singular vectors and values via its singular value decomposition (SVD):
\begin{equation}
\res = \sum_\sigma s_\sigma \mathbf{u}_\sigma \mathbf{v}_\sigma^\dagger \, .
\end{equation}
where $s_\sigma$ are the singular values and $\mathbf{u}_\sigma$ and $\mathbf{v}_\sigma$ are the corresponding left and right singular vectors, respectively.
For nondegenerate QNMs, the SVD only has one non-zero singular value $s_0$ and corresponding left and right singular vectors $\mathbf{u}_0$ and $\mathbf{v}_0$:
\begin{equation}
\label{eq-residue-svd}
\res = s_0 \mathbf{u}_0 \mathbf{v}_0^\dagger \, .
\end{equation}

\textbf{Reciprocity Relations between Left and Right Singular Vectors:}
An object described by a $\mathbf{T}$-matrix in terms of VSWs is reciprocal if \citep{Asadova_2025}:
\begin{equation}
T_{\ell m \lambda \mu}^{ij} = (-1)^{m +\mu}  T_{\ell (-m) \lambda (-\mu)}^{ji}\, ,
\end{equation}
where $i,j$ indicate the polarization of the VSW. $\ell$ and $m$ are the degree and order of the scattered multipole, while $\lambda$ and $\mu$ are the degree and order of the incident multipole (not to be confused with wavelength and permeability). This condition for reciprocity can be equivalently expressed as \citep{Mishchenko1996T}:
\begin{equation}
\label{eq-reciprocityT}
\mathbf{T} = \mathbf{P} \mathbf{T}^\mathsf{T} \mathbf{P}\, ,
\end{equation}
where $\mathbf{P}$ is the signed permutation:
\begin{equation}
\mathbf{P}_{\ell,m,\ell^\prime, m^\prime}^{ij} = (-1)^{m} \delta_{\ell,\ell^\prime} \delta_{m,-m^\prime} \delta_{i,j}\, .
\end{equation}

$\mathbf{P}$ is unitary and self-inverse:
\begin{equation}
\mathbf{P}^{-1} = \mathbf{P}^\dagger = \mathbf{P}^\mathsf{T} = \mathbf{P} \, .
\end{equation}

Using the properties of $\mathbf{P}$, we reformulate Eq. (\ref{eq-reciprocityT}) as:
\begin{equation}
\underbrace{\mathbf{P}\mathbf{T}}_\mathbf{F} =  \underbrace{\mathbf{T}^\mathsf{T} \mathbf{P}^\mathsf{T}}_{\mathbf{F}^\mathsf{T}} \, .
\end{equation}
Here, we have introduced $\mathbf{F}$, which is complex-symmetric. As such, $\mathbf{F}$ has a Takagi factorization \citep{Chebotarev_2014}:
\begin{equation}
\mathbf{F} = \mathbf{W} \boldsymbol{\Sigma} \mathbf{W}^\mathsf{T}\, ,
\end{equation}
where $\mathbf{W}$ is unitary and $\boldsymbol{\Sigma}$ is diagonal with non-negative real entries.
Using the Takagi factorization, we can express a valid SVD of $\mathbf{T}$ as:
\begin{equation}
\mathbf{T} = \underbrace{\mathbf{P}\mathbf{W}}_{\boldsymbol{\Xi} \mathbf{U}} \boldsymbol{\Sigma} \underbrace{\mathbf{W}^\mathsf{T}}_{\mathbf{V}^\dagger \boldsymbol{\Xi}^\dagger} \, .
\end{equation}
Here, $\boldsymbol{\Xi}$ is a unitary diagonal matrix, which accounts for the phase freedom in the SVD. From this expression, we identify the left and right singular vectors as:
\begin{equation}
\mathbf{P} \xi_i \mathbf{u}_i = \xi_i^* \mathbf{v}_i^* \, .
\end{equation}
Here, $\xi_i$ is the $i$-th diagonal element of $\boldsymbol{\Xi}$. In the following derivation, we fix the phase freedom of the SVD by choosing $\xi_i = 1$ for all $i$:
\begin{equation}
\label{eq-reciprocity-relation}
\mathbf{v}_i^* = \mathbf{P} \mathbf{u}_i \, .
\end{equation}

As we have seen, this relation of left and right singular vectors of $\mathbf{T}$ is a direct consequence of reciprocity. To hold for any $\omega$, it also has to hold for the residues $\res$ of the pole-expansion.

For degenerate singular values, one has to be a bit more careful, as the singular vectors can be chosen arbitrarily within the degenerate subspace. However, one can always find an SVD fulfilling the above relation (\textit{i.e.}, via the Takagi factorization as shown).

\textbf{Further Simplifications:}
Using Eq. (\ref{eq-reciprocity-relation}) we can express the residue from Eq. (\ref{eq-residue-svd}) as:
\begin{equation}
\res = s_0 \mathbf{u}_0 (\mathbf{P} \mathbf{u}_0)^\mathsf{T} \, ,
\end{equation}

which we can plug into the expression for $\mathbf{q}_m$:
\begin{equation}
\mathbf{q}_m = s_0 \mathbf{u}_0 \frac{ \mathbf{u}_0^\mathsf{T} \mathbf{P} \mathbf{C}_{-\mathbf{r}_\mathrm{p}}^{(3)} \mathbf{R} \mathbf{d} }{ \sqrt{- \tilde{\omega}_m  \frac{\mathrm{i} \varepsilon_0}{k_0^3} \mathbf{d}^\mathsf{T} \mathbf{R}^{-1} \mathbf{C}_{\mathbf{r}_\mathrm{p}}^{(3)} s_0 \mathbf{u}_0 \mathbf{u}_0^\mathsf{T} \mathbf{P} \mathbf{C}_{-\mathbf{r}_\mathrm{p}}^{(3)} \mathbf{R} \mathbf{d} }}\, .
\end{equation}

By pulling out the scalar terms, we get:
\begin{equation}
\label{eq-frac1}
\mathbf{q}_m = \underbrace{\sqrt{\frac{\mathrm{i} s_0 k_0^3}{\tilde{\omega}_m \varepsilon_0}}}_{\normalsize \tilde \omega_m \sqrt{\frac{\mathrm{i} s_0}{\varepsilon_0 c_0^3}}} \mathbf{u}_0 \frac{\highlight{cyan}{\mathbf{u}_0^\mathsf{T} \mathbf{P} \mathbf{C}_{-\mathbf{r}_\mathrm{p}}^{(3)} \mathbf{R} \mathbf{d} }}{ \sqrt{ \highlight{orange}{ \phantom{\Big(} \mathbf{d}^\mathsf{T} \mathbf{R}^{-1} \mathbf{C}_{\mathbf{r}_\mathrm{p}}^{(3)}  \mathbf{u}_0} \highlight{cyan}{ \mathbf{u}_0^\mathsf{T} \mathbf{P} \mathbf{C}_{-\mathbf{r}_\mathrm{p}}^{(3)} \mathbf{R} \mathbf{d} \phantom{\Big)} }}}\, .
\end{equation}

\textbf{Evaluating the Fraction in Eq. (\ref{eq-frac1}):}
Intuitively, reciprocity tells us that the last fraction equals 1. In essence, the term highlighted in orange results when reversing the source and `receiver' (the field distribution we project onto) of the blue term. By reciprocity, these two terms are equal, leaving us with a sign ambiguity, which is, anyway, inherent to the QNM-fields.

To eliminate the fraction more rigorously, we use the following relations:
\begin{equation}
\begin{aligned}
\mathbf{C}_{\mathbf{r}_\mathrm{p}}^{(3)} &= \mathbf{P} \mathbf{C}_{-\mathbf{r}_\mathrm{p}}^{(3)\mathsf{T}} \mathbf{P},\\
\mathbf{R}^{-1} &= \mathbf{P}\mathbf{R}^\mathsf{T}\mathbf{P},\\
\mathbf{P} \mathbf{d} &= \mathbf{d}\, .
\end{aligned}
\end{equation}

The last relations directly follows from the structure of $\mathbf{d}$ (with only one non-zero entry at $\ell=1, m=0$) which is unaffected by $\mathbf{P}$. The other two relations are derived in Section~\ref{symmetry-C} from the translation addition theorems and in Section~\ref{symmetry-R} from the symmetries of the Wigner D-functions for $\mathbf{C}_{\mathbf{r_p}}^{(3)}$ and $\mathbf{R}^{ -1}$, respectively.

Using these relations and considering that the transpose of a scalar is the scalar itself, we find:
\begin{equation}
\begin{aligned}
\mathbf{u}_0^\mathsf{T} \mathbf{P} \mathbf{C}_{-\mathbf{r}_\mathrm{p}}^{(3)} \mathbf{R} \mathbf{d}
    & =  \mathbf{u}_0^\mathsf{T} \mathbf{P} \mathbf{C}_{-\mathbf{r}_\mathrm{p}}^{(3)} \mathbf{R} \mathbf{P} \mathbf{d} \\
    & =  \mathbf{d}^\mathsf{T} \mathbf{P}\mathbf{R}^\mathsf{T} \mathbf{C}_{-\mathbf{r}_\mathrm{p}}^{(3)\mathsf{T}} \mathbf{P} \mathbf{u}_0 \\
    & =  \mathbf{d}^\mathsf{T} \mathbf{P}\mathbf{R}^\mathsf{T} \mathbf{P} \mathbf{P} \mathbf{C}_{-\mathbf{r}_\mathrm{p}}^{(3)\mathsf{T}} \mathbf{P} \mathbf{u}_0 \\
    & =  \mathbf{d}^\mathsf{T} \mathbf{R}^{-1} \mathbf{C}_{\mathbf{r}_\mathrm{p}}^{(3)} \mathbf{u}_0 \, .
\end{aligned}
\end{equation}
Thus, we find that the fraction in Eq. (\ref{eq-frac1}) equals 1.

\begin{framed}
\textbf{Normalized QNM in VSWs}\\
\begin{equation}
\mathbf{q}_m = \tilde \omega_m \sqrt{\frac{\mathrm{i} s_0}{\varepsilon_0 c_0^3}} \mathbf{u}_0\, .
\end{equation}

Here, $\mathbf{q}_m$ are the normalized QNM fields expressed as coefficient vectors in terms of outgoing VSWs, $\tilde \omega_m$ is the complex resonance frequency of the QNM, $s_0$ is the non-zero singular value of the residue $\res$ corresponding to the QNM, and $\mathbf{u}_0$ is the corresponding left singular vector, with the phase-freedom fixed by the Takagi-factorization of $\mathbf{F}=\mathbf{P}\res$.
\end{framed}

We have found the normalized QNMs directly from the residues in the pole-expansion of the T-matrix.
This is quite convenient, as we can avoid numerically evaluating any integrals or limits.
Note that the fields can only be evaluated outside the scatterer, as the expansion of the scattered field in singular VSWs is valid only there.

\balancecolsandclearpage
\onecolumngrid
\subsection{Numerical Verification}

\begin{figure*}[!htbp]
\centering
\includegraphics[width=1\linewidth]{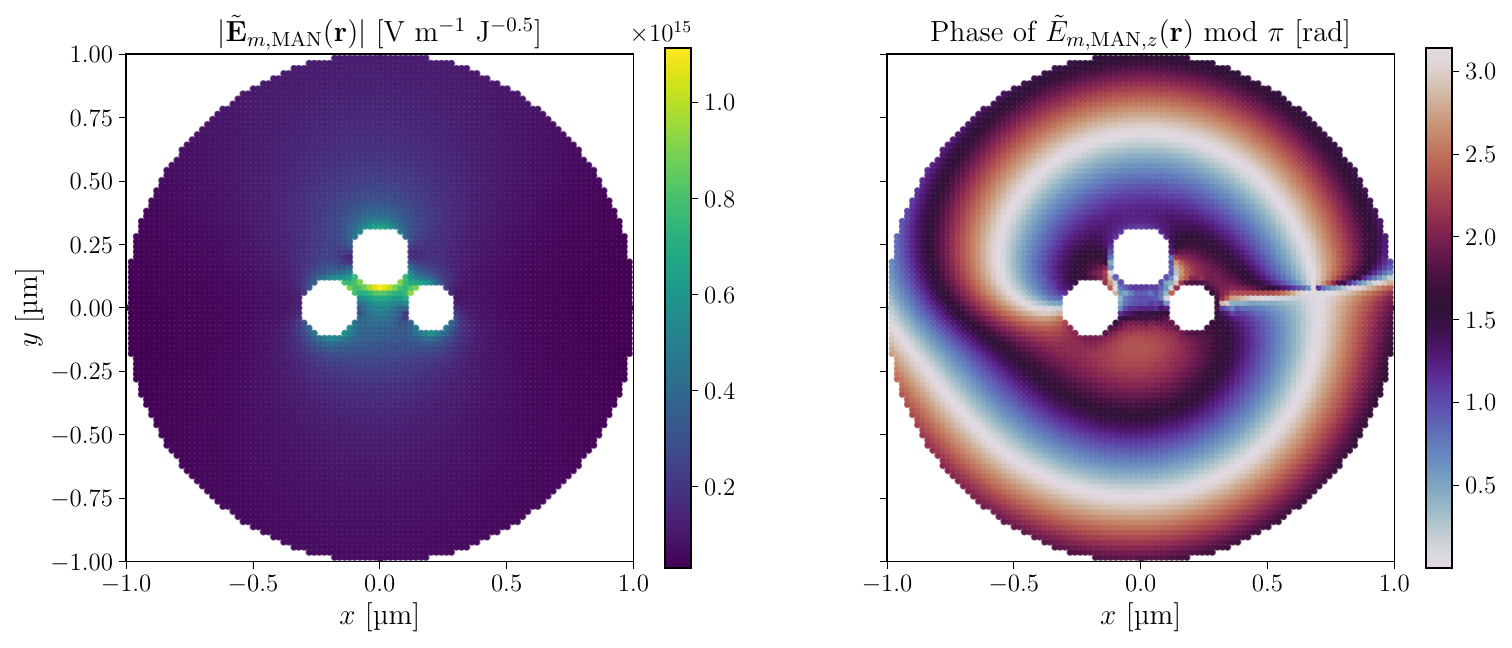}
\caption[]{Magnitude and phase of the normalized QNM obtained via MAN. The field is not shown within the geometry (a tetrahedron of spheres) as \texttt{treams} does not allow for evaluating the field there. The results from \texttt{treams} are not shown as they are visually indistinguishable.}
\label{fig-MAN}
\end{figure*}

\begin{figure*}[!htbp]
\centering
\includegraphics[width=1\linewidth]{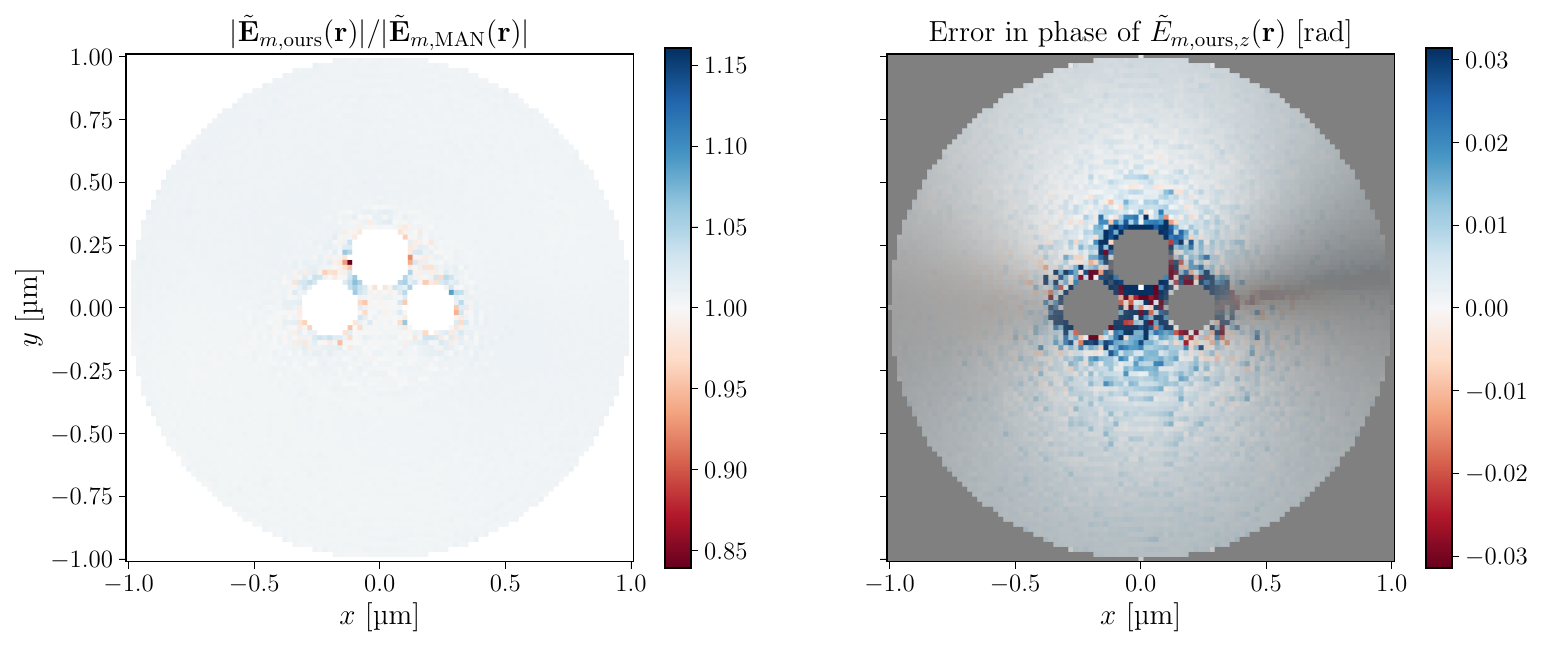}
\caption[]{The magnitude and phase resulting from the proposed solution are compared to the reference solutions from MAN. In the phase-error regions where the considered field component is close to zero, they are masked out.}
\label{fig-Error}
\end{figure*}

To verify the developed normalization scheme, we compare the QNMs obtained from our scheme to QNMs obtained from MAN \citep{Wu_2023} (see Figure~\ref{fig-MAN}) with a volume integral over the complete simulation domain including PMLs. We find very good agreement between the two methods (see Figure~\ref{fig-Error}). The agreement also holds when considering T-matrices in a local basis as commonly used in \texttt{treams} (as shown in Figure~\ref{fig-MAN} and Figure~\ref{fig-Error}).

\clearpage
\onecolumngrid
\subsection{Symmetry Relations of Translation and Rotation Operators}

\subsubsection{Proof $\mathbf{C}_{\rp}^{(3)} = \mathbf{P} \mathbf{C}_{ -\rp}^{(3)\mathsf{T}} \mathbf{P}$}\label{symmetry-C}

The translational addition theorems \citep{cruzan1962translational} are the backbone of T-matrix based multiple scattering calculations. Here we use the relation between singular VSWs $\Plms{3}{}$ centered at the origin and regular VSWs $\Ps{\lam \mu \pol}{1}{ -\rp}$ centered at point $\rp$ (Eq. (6a) in supplements to Ref. \citep{beutel_efficient_2021}; we have swapped the positions of singular and regular VSWs for a more convenient notation). We use the index $s$ instead of $p$ in the main text to indicate polarization:
\begin{equation}
\begin{aligned}
\Ps{\ell m \pol}{3}{} = \sum_{\ell=1}^\infty \sum_{m=-\ell}^{\ell} \bigg[A_{\lam\mu \ell m} (\rp,k) \Ps{\lam \mu \pol}{1}{-\rp} + B_{\lam\mu \ell m} (\rp,k) \Ps{\lam \mu (-\pol)}{1}{-\rp} \bigg] \, .
\end{aligned}
\end{equation}

We start from the expressions for the translation coefficients as they are implemented in \texttt{treams}. Eq. (C.1) in the appendix of Ref. \citep{beutelTreamsTmatrixbasedScattering2024} provides expressions for the translation coefficients $A_{\lam\mu \ell m}(\rp,k)$ and $B_{\lam\mu \ell m}(\rp,k)$. In an attempt to make the used expressions more digestable we disect them into three parts each. Further, we have replaced $p$ and $q$ from references \citep{beutel_efficient_2021, beutelTreamsTmatrixbasedScattering2024} with the symbols $\varsigma$ and $\zeta$, respectively, to avoid a clash of notation.
\begin{equation}
\label{eq-A}
\begin{aligned}
A_{\lam\mu \ell m} (\rp,k) = b_{\lam \mu \ell m} (\rp) \hspace{-0.8cm} \sum_{\varsigma=\max(|\ell-\lam|,|m-\mu|)}^{\ell+\lam} \hspace{-0.8cm} a(\ell,m,\lam, -\mu, \varsigma, \varsigma) \underbrace{[\ell(\ell+1)+\lam(\lam+1)-\varsigma(\varsigma+1)]z_{\varsigma}^{(n)}(kr)P_{\varsigma}^{m-\mu}(\cos\theta)}_{f_{\lam \mu \ell m \varsigma}(\rp, k)} \, ,
\end{aligned}
\end{equation}
where the sum only runs over even $\ell + \lam + \varsigma$. And
\begin{equation}
\label{eq-B}
\begin{aligned}
B_{\lam\mu \ell m} (\rp,k) = b_{\lam \mu \ell m} (\rp) \hspace{-0.8cm} \sum_{\varsigma=\max(|\ell-\lam|,|m-\mu|)}^{\ell+\lam} \hspace{-0.8cm} a(\ell,m,\lam, -\mu, \varsigma, \varsigma-1) \underbrace{\sqrt{((\ell+\lam+1)^2-\varsigma^2)(\varsigma^2-(\ell-\lam)^2)} z_{\varsigma}^{(n)}(kr)P_{\varsigma}^{m-\mu}(\cos\theta)}_{g_{\lam \mu \ell m \varsigma}(\rp, k)} \, ,
\end{aligned}
\end{equation}
where odd $\ell + \lam + \varsigma$ are considered in the sum. The function $a(\ell,m,\lam,\mu, \varsigma, \zeta)$ is defined as:
\begin{equation}
\label{eq-treams}
\begin{aligned}
a(\ell,m,&\lam,\mu, \varsigma, \zeta)
= (2\varsigma + 1) \ii^{\lam - \ell + \varsigma} \sqrt{\frac{(\varsigma-m-\mu)!}{(\varsigma+m+\mu)!}} 
\begin{pmatrix}
\ell & \lam & \varsigma \\
m & \mu & -(m+\mu)
\end{pmatrix}
\begin{pmatrix}
\ell & \lam & \zeta \\
0 & 0 & 0
\end{pmatrix} \, .
\end{aligned}
\end{equation}
Here the bracketed expressions on the right are the Wigner-$3jm$ symbols \citep{Varshalovich_1988} and should not be mistaken for matrices. The prefactor $b_{\lam \mu \ell m} (\rp)$ is defined as:
\begin{equation}
b_{\lam \mu \ell m} (\rp) = \frac{1}{2} (-1)^m \sqrt{\frac{(2\ell+1)(2\lam +1)}{\ell(\ell+1)\lam(\lam+1)}} e^{\ii (m-\mu) \varphi} \, .
\end{equation}

\paragraph{Symmetries of the Subexpressions}

Throughout this derivation we'll repeatedly use the fact that all indices $\nu = \lam,\mu,\ell,m,\varsigma, \zeta \in \mathbb{Z}$. As such $( -1)^{ -\nu} = ( -1)^{\nu}$. We deduce the symmetries of the subexpressions directly:
\begin{equation}
\begin{aligned}
b_{\lam \mu \ell m} (\rp) &= \frac{1}{2} (-1)^m \sqrt{\frac{(2\ell+1)(2\lam +1)}{\ell(\ell+1)\lam(\lam+1)}} e^{\ii (m-\mu) \varphi} \, ,\\
b_{\ell (-m) \lam (-\mu)} (\rp) &=  \frac{1}{2} (-1)^{-\mu} \sqrt{\frac{(2\ell+1)(2\lam +1)}{\ell(\ell+1)\lam(\lam+1)}} e^{\ii (-\mu + m) \varphi} = b_{\lam \mu \ell m} (\rp) (-1)^{m-\mu} \, ,\\
b_{\lam \mu \ell m} (-\rp) &= \frac{1}{2} (-1)^m \sqrt{\frac{(2\ell+1)(2\lam +1)}{\ell(\ell+1)\lam(\lam+1)}} e^{\ii (m-\mu) (\varphi + \pi)} = b_{\lam \mu \ell m} (\rp) (-1)^{m-\mu} \, .
\end{aligned}
\end{equation}
And
\begin{equation}
\begin{aligned}
f_{\lam \mu \ell m \varsigma}(\rp, k) &= [\ell(\ell+1)+\lam(\lam+1)-\varsigma(\varsigma+1)]z_{\varsigma}^{(n)}(kr)P_{\varsigma}^{m-\mu}(\cos\theta) \, , \\
f_{\ell (-m) \lam (-\mu) \varsigma}(\rp, k) &= [\lam(\lam+1)+\ell(\ell+1)-\varsigma(\varsigma+1)]z_{\varsigma}^{(n)}(kr)P_{\varsigma}^{-\mu+m}(\cos\theta)\\ &= f_{\lam \mu \ell m \varsigma}(\rp, k)\, , \\
f_{\lam \mu \ell m \varsigma}(-\rp, k) &= [\ell(\ell+1)+\lam(\lam+1)-\varsigma(\varsigma+1)]z_{\varsigma}^{(n)}(kr)\underbrace{P_{\varsigma}^{m-\mu}(-\cos\theta)}_{(-1)^{\varsigma+\mu-m}P_{\varsigma}^{m-\mu}(\cos\theta)}\\ &= (-1)^{\varsigma+\mu-m} f_{\lam \mu \ell m \varsigma}(\rp, k) \, .
\end{aligned}
\end{equation}

Analogously,
\begin{equation}
\begin{aligned}
g_{\ell (-m) \lam (-\mu) \varsigma}(\rp, k) &= g_{\lam \mu \ell m \varsigma}(\rp, k)\, , \\
g_{\lam \mu \ell m \varsigma}(-\rp, k) &= (-1)^{\varsigma+\mu-m} g_{\lam \mu \ell m \varsigma}(\rp, k) \, .
\end{aligned}
\end{equation}

And lastly for $a(\ell,m,\lam,\mu, \varsigma, \zeta)$ we need to pay special attention, as we give $\mu$ a sign flip in Eqs. (\ref{eq-A}) and (\ref{eq-B}):
\begin{equation}
\begin{aligned}
a(\ell,m,\lam,-\mu,\varsigma,\zeta) &= S a(\lam,-\mu,\ell,m,\varsigma,\zeta) \\
\iff \cancel{(2\varsigma+1)}\ii^{\lam-\ell+ \cancel{\varsigma}} & \cancel{\sqrt{\frac{(\varsigma-m+\mu)!}{(\varsigma+m-\mu)!}}} 
\begin{pmatrix}
\ell & \lam & \varsigma \\
m & \mu & -(m+\mu)
\end{pmatrix}
\begin{pmatrix}
\ell & \lam & \zeta \\
0 & 0 & 0
\end{pmatrix} = \\
S \cancel{(2\varsigma+1)}\ii^{\ell-\lam+ \cancel{\varsigma}} & \cancel{\sqrt{\frac{(\varsigma+\mu-m)!}{(\varsigma-\mu+m)!}}} 
\begin{pmatrix}
\lam & \ell & \varsigma \\
\mu & m & -(m+\mu)
\end{pmatrix}
\begin{pmatrix}
\lam & \ell & \zeta \\
0 & 0 & 0
\end{pmatrix}\\
\end{aligned}
\end{equation}

Using the symmetries of the Wigner-$3jm$ symbols \citep{Varshalovich_1988}:

\begin{equation}
\begin{aligned}
\iff & & (-1)^{\lam-\ell}
\begin{pmatrix}
\ell & \lam & \varsigma \\
m & \mu & -(m+\mu)
\end{pmatrix}
\begin{pmatrix}
\ell & \lam & \zeta \\
0 & 0 & 0
\end{pmatrix} &= 
S \begin{pmatrix}
\ell & \lam & \varsigma\\
m & \mu & -(m+\mu)
\end{pmatrix} (-1)^{\ell+\lam+\varsigma}
\begin{pmatrix}
\ell & \lam & \zeta \\
0 & 0 & 0
\end{pmatrix} (-1)^{\ell+\lam+\zeta}\\
\iff & & S &= (-1)^{\lam-\ell}(-1)^{\varsigma+\zeta}
\end{aligned}
\end{equation}

We conclude
\begin{equation}
a(\ell,m,\lam,-\mu,\varsigma,\zeta) = (-1)^{\lam-\ell}(-1)^{\varsigma+\zeta} a(\lam,-\mu,\ell,m,\varsigma,\zeta)
\end{equation}

\paragraph{Symmetries of the Translation Coefficients}

For readability we'll drop the limits of the sum. They are assumed to be the same as in (\ref{eq-treams}) and above. To express the fact that $\ell + \lam + \varsigma$ is even we introduce $c \in \mathbb{Z}$ such that $\ell+\lam+\varsigma = 2c$.
\begin{equation}
\begin{aligned}
A_{\lam\mu \ell m}(\rp,k) &= b_{\lam \mu \ell m} (\rp) \sum_{\varsigma} a(\ell,m,\lam, -\mu, \varsigma, \varsigma) f_{\lam \mu \ell m \varsigma}(\rp, k) \, , \\
A_{\ell (-m) \lam (-\mu)}(\rp,k) &= (-1)^{m-\mu} b_{\lam \mu \ell m} (\rp) \sum_{\varsigma} a(\ell,m,\lam, -\mu, \varsigma, \varsigma) (-1)^{\lam-\ell}\cancel{(-1)^{\varsigma+\varsigma}}f_{\lam \mu \ell m \varsigma}(\rp, k) \\
&= (-1)^{m+\mu} (-1)^{\ell+\lam} A_{\lam\mu \ell m}(\rp,k) \, , \\
A_{\lam\mu \ell m}(-\rp,k) &= b_{\lam \mu \ell m}(\rp) (-1)^{m-\mu} \sum_{\varsigma} a(\ell,m,\lam, -\mu, \varsigma, \varsigma) f_{\lam \mu \ell m \varsigma}(\rp, k) (-1)^{\cancel{2c}-\ell-\lam} (-1)^{\mu - m}\\
&= (-1)^{\ell+\lam+1} A_{\lam\mu \ell m}(\rp,k) = (-1)^{m-\mu} A_{\ell (-m) \lam (-\mu)}(\rp,k)
\end{aligned}
\end{equation}

Here $\ell + \lam + \varsigma$ is odd we introduce $c \in \mathbb{Z}$ such that $\ell+\lam+\varsigma = 2c + 1$
\begin{equation}
\begin{aligned}
B_{\lam\mu \ell m}(\rp,k) &= b_{\lam \mu \ell m} (\rp) \sum_{\varsigma} a(\ell,m,\lam, -\mu, \varsigma, \varsigma-1) g_{\lam \mu \ell m \varsigma}(\rp, k) \, , \\
B_{\ell (-m) \lam (-\mu)}(\rp,k) &= (-1)^{m-\mu} b_{\lam \mu \ell m} (\rp) \sum_{\varsigma} a(\ell,m,\lam, -\mu, \varsigma, \varsigma-1) (-1)^{\lam-\ell}(-1)^{\cancel{\varsigma+\varsigma}-1}g_{\lam \mu \ell m \varsigma}(\rp, k) \\
&= (-1)^{m+\mu} (-1)^{\ell+\lam+1} B_{\lam\mu \ell m}(\rp,k) \, , \\
B_{\lam\mu \ell m}(-\rp,k) &= b_{\lam \mu \ell m}(\rp) (-1)^{m-\mu} \sum_{\varsigma} a(\ell,m,\lam, -\mu, \varsigma, \varsigma-1) f_{\lam \mu \ell m \varsigma}(\rp, k) (-1)^{\cancel{2c}+1-\ell-\lam} (-1)^{\mu - m} \\
&= (-1)^{\ell+\lam} B_{\lam\mu \ell m}(\rp,k) = (-1)^{m-\mu} B_{\ell (-m) \lam (-\mu)}(\rp,k)
\end{aligned}
\end{equation}

\vspace{6cm}
\twocolumngrid
In summary
\begin{equation}
\begin{aligned}
A_{\lam\mu \ell m}(-\rp,k) &= (-1)^{m+\mu} A_{\ell(-m) \lam (-\mu)}(\rp,k) \, , \\
B_{\lam\mu \ell m}(-\rp,k) &= (-1)^{m+\mu} B_{\ell(-m) \lam (-\mu)}(\rp,k) \, .
\end{aligned}
\end{equation}
These translation coefficients $A_{\lam\mu \ell m}(\rp,k)$ and $B_{\lam\mu \ell m}(\rp,k)$ are the matrix elements of the operator $\mathbf{C}_{\rp}^{(3)}$ \citep{beutel_efficient_2021}. Therefore,
\begin{equation}
\mathbf{C}_{\rp}^{(3)} = \mathbf{P} \mathbf{C}_{-\rp}^{(3)\mathsf{T}} \mathbf{P} \, .
\end{equation}

\subsubsection{Proof $\mathbf{R}^{ -1} = \mathbf{P}\mathbf{R}^\mathsf{T}\mathbf{P}$}\label{symmetry-R}

For notational convenience we show the equivalent relation for $\mathbf{R} = \mathbf{R^\prime}^{ -1}$:
\begin{equation}
\mathbf{R} = \mathbf{P}(\mathbf{R}^{-1})^\mathsf{T}\mathbf{P},
\end{equation}
In the following, we specify the rotation axis and degree in terms of zyz-Euler angles. Whenever no rotation angles are specified we assume $\mathbf{R} = \mathbf{R}(\alpha, \beta, \gamma)$. In the zyz-Euler convention $\mathbf{R}( -\gamma, -\beta, -\alpha) = \mathbf{R}^{ -1}(\alpha, \beta, \gamma)$. The matrix elements are given by the Wigner D-functions \citep{Varshalovich_1988}:

\begin{equation}
\bra{\ell m} \mathbf{R}(\alpha, \beta, \gamma) \ket{\ell\mprime} = D_{m \mprime}^\ell(\alpha, \beta, \gamma) \, ,
\end{equation}
where multipoles of different degree $\ell$ are not mixed.
Accordingly the matrix element of the inverse reads as:

\begin{equation}
\bra{\ell m} \mathbf{R}(-\gamma, -\beta, -\alpha) \ket{\ell\mprime} = D_{m \mprime}^\ell(-\gamma, -\beta, -\alpha) \, .
\end{equation}

The matrix element of $(\mathbf{R}^{ -1})^\mathsf{T}$ thus are:

\begin{equation}
\begin{aligned}
M:&= \bra{\ell m} \mathbf{R}^\mathsf{T}(-\gamma, -\beta, -\alpha)\ket{\ell\mprime} \\&= D_{\mprime m}^\ell(-\gamma, -\beta, -\alpha) \, , 
\end{aligned}
\end{equation}
where the column and row indixes have been switched.
Using $\mathbf{P}\mathbf{P} = \mathbf{I}$ and the fact that P flips the sign of $m$ and applies a $( -1)^m$ sign flip we get:
\begin{equation}
\begin{aligned}
M &= \bra{\ell m} \mathbf{P}\mathbf{P} \mathbf{R}^\mathsf{T}(-\gamma, -\beta, -\alpha) \mathbf{P}\mathbf{P}\ket{\ell\mprime} \\
&= \underbrace{(-1)^{\mprime-m}}_\varepsilon \bra{\ell(-m)} \mathbf{P} \mathbf{R}^\mathsf{T}(-\gamma, -\beta, -\alpha) \mathbf{P}\ket{\ell-\mprime} \, .
\end{aligned}
\end{equation}
Further, we know from \citep{Varshalovich_1988} that:
\begin{equation}
\begin{aligned}
M &= D_{\mprime m}^\ell(-\gamma, -\beta, -\alpha) = \varepsilon D_{- m - \mprime}^\ell(\alpha, \beta, \gamma) \\
&= \varepsilon \bra{\ell(-m)} \mathbf{R}(\alpha, \beta, \gamma) \ket{\ell-\mprime}
\end{aligned}
\end{equation}
As this is true for any matrix element, we know that
\begin{equation}
\mathbf{P} \mathbf{R}^\mathsf{T}(-\gamma, -\beta, -\alpha) \mathbf{P} = \mathbf{R}(\alpha, \beta, \gamma)
\end{equation}

\section{FEM Convergence Study}


Let us leverage the semi-analytical solvability of the tetrahedron example to study the convergence of FEM evaluations of $\mathbf{T}(k)$. The convergence analysis emphasizes that the accuracy of the underlying numerical solution imposes a lower bound on the error of the pole-expansion with respect to the physical system.

\begin{figure}[!htbp]
\centering
\includegraphics[width=0.65\linewidth]{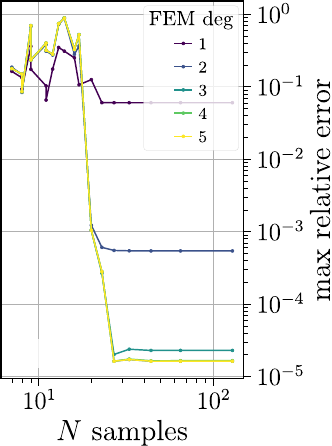}
\caption[]{Convergence of the pole-expansion of the T-matrix of a tetrahedron of spheres as was considered in Figure 2 of the main text. In contrast to Figure 2, the T-matrix samples were evaluated here using a finite element method (FEM) as implemented in JCMsuite. The mesh and PML were fixed across different $k$ to avoid discretization noise between samples. The finite element degree was varied to examine the effect of FEM solution quality on the convergence of the AAA-based pole-expansion (differently colored lines). All other aspects are identical to Figure 2.}
\label{fig-FEM_tetrahedron_aaa}
\end{figure}

In Figure~\ref{fig-FEM_tetrahedron_aaa}, we study the convergence of the pole-expansion of the T-matrix of the tetrahedron of spheres considered in Figure 2 of the main text when using a finite element method (FEM) to evaluate the T-matrix samples instead of the semi-analytical multiple-scattering solution. Across different $k$, the finite element discretization was fixed by setting the mesh and PML parameters. The relative error [Eq. (14) of the main text] is evaluated with respect to the semi-analytical reference solutions. In comparison to the results shown Figure 2 of the main text, the error plateaus at a significantly higher level, which depends strongly on the finite element degree (increasing from dark blue to yellow - see legend). These plateaus indicate that the pole-expansion converges further, and the final error is dominated by the mismatch between FEM evaluations and the reference solutions. We note that even at the highest FEM accuracy, the plateau is reached after less than 40 samples, highlighting the efficiency of \texttt{tensorAAA} for constructing accurate T-matrix representations. Between a finite element degree of 4 and 5, no significant improvement of the final accuracy is observed, indicating that other numerical aspects (\textit{e.g.}, mesh) might become limiting factors here. We further note that of the found poles, only the dominant ones are in good agreement to the poles shown in Figure 2 of the main text, which is easily understood when considering that many of the poles (in particular the peripheral and low $Q$ ones) serve to express relatively small details in the scattering response, which differ here due to the discretization error.

\section{Approximating Individual Matrix Elements Separately}

\begin{figure}[!htbp]
\centering
\includegraphics[width=0.85\linewidth]{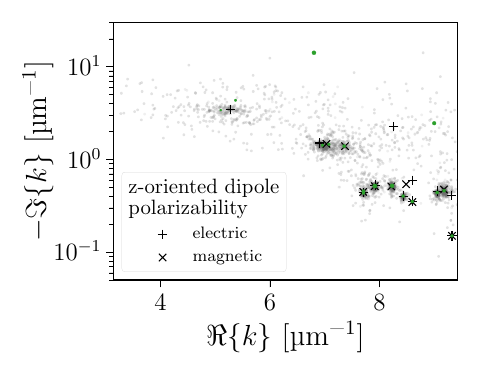}
\caption[]{Poles of the rational approximant to the T-matrix constructed by \texttt{tensorAAA} [green dots; matches with Figure 2(b) of the main text] and of approximants to the individual matrix elements constructed with the standard scalar-valued AAA-algorithm (gray dots; the T-matrix elements corresponding to the z-oriented magnetic and electric dipole polarizabilities are indicated by plus and cross markers, respectively.).}
\label{fig-individual-poles}
\end{figure}

In this supplement we compare the poles found by applying \texttt{tensorAAA} to all entries of the T-matrix -- truncated by an $\ell_\mathrm{max} = 3$ -- simultaneously (as was demonstrated in the main text) to applying the standard AAA-algorithm to each element of the T-matrix separately. As shown in Figure~\ref{fig-individual-poles} this second approach yields shifted poles per element.

\section{Fluctuations in the Convergence of the \texttt{tensorAAA} Pole-Expansion}


\begin{figure}[!htbp]
\centering
\includegraphics[width=0.8\linewidth]{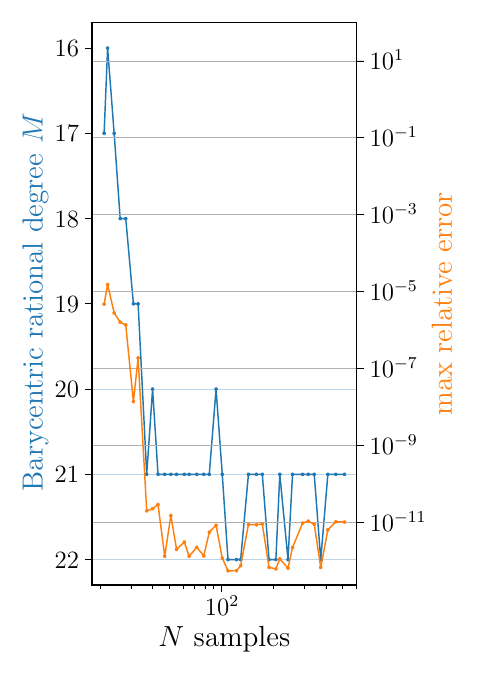}
\caption[]{Joint plot of the AAA approximation degree and the corresponding relative error for the tetrahedron example with $\text{tol}=10^{ -6}$ and varying number of samples. The relative error, also shown in Figure 2(c) of the main text, exhibits some discrete jumps in the convergence plateau, in addition to the typical random fluctuations.}
\label{fig-fluctuations-convergence}
\end{figure}

When using the \texttt{tensorAAA} algorithm to find a pole-expansion of the T-matrix elements, we observe rapid convergence towards a low relative error, until a plateau is reached, as shown in Figure 2(c) of the main text. However, within this plateau, we observe some discrete jumps in the relative error, which are particularly pronounced for $\text{tol}=10^{ -6}$ in the tetrahedron example. To identify the cause of these jumps, we jointly plot the degree of the AAA approximation and the corresponding relative error in Figure~\ref{fig-fluctuations-convergence}. It is evident that the jumps in the relative error coincide with jumps in the degree of the AAA approximation. The degree of the AAA approximation is determined adaptively by continuing the iteration until an acceptable residual error is met. Small fluctuations in the residual error in the last iterations can thus lead to these discrete jumps in the relative error, as the degree might or might not be increased by one.

\section{Dipolar Condition for Coincident BICs}

\begin{figure*}[!htbp]
\centering
\includegraphics[width=0.6\linewidth]{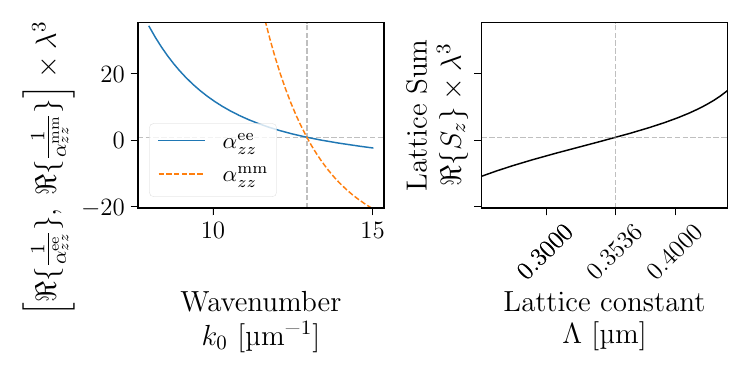}
\caption[]{Ref. \citep{evlyukhin_polarization_2021} proposed different procedures to obtain coincident BICs for lattices of small (i.e. dipolar) meta-atoms. Using procedure (i) we find the condition for coincident electric and magnetic dipolar BICs by identifying the wavenumber for which the real parts of the (inverse) z-oriented electric and magnetic polarizabilities of the isolated meta-atom match (left). The imaginary parts do not need to be considered, as the optical theorem already enforces their equality. Next, the dipolar lattice sum $S_z(\Lambda)$ is evaluated for a range of lattice constants $\Lambda$. The condition for coincident BICs in this dipolar model is then given by the intersection of the real part of the inverse polarizability with the real part of the lattice sum (right). In the main text, we denote the corresponding lattice constant as $\Lambda_\mathrm{dipole}$}
\label{fig-dipolar_BIC}
\end{figure*}

In Figure~\ref{fig-dipolar_BIC}, we illustrate the procedure to identify the dipolar condition for coincident electric and magnetic dipolar BICs as proposed by \cite{evlyukhin_polarization_2021}. The procedure to find the wavenumber and lattice constant for which coincident dipolar BICs occur is straightforward when considering the meta-atom to be predominantly dipolar. Then, no multipolar contributions other than the z-oriented dipoles are permitted, which also becomes clear from the group theoretical considerations that lead to selection rules for the multipolar contents of $\Gamma$-point BICs \citep{sadrieva_multipolar_2019}. However, these considerations also indicate that, beyond the dipolar approximation, additional multipoles are permitted, leading to a more sophisticated multiple-scattering resonance condition that no longer admits a simple graphical solution. How the BICs can still be found while still leveraging the power of lattice sums is discussed in the main text.

\clearpage
\section{Restoring xy-Plane Mirror Symmetry}

\begin{figure}[!htbp]
\centering
\includegraphics[width=0.85\linewidth]{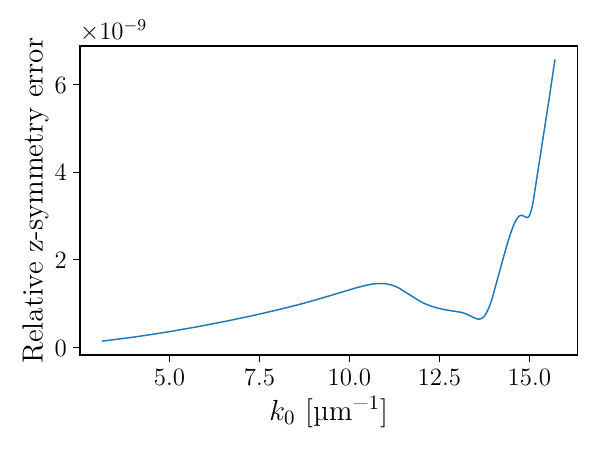}
\caption[]{Spectrally resolved relative error of the T-matrix of the cylinder example due to broken $\boldsymbol{\sigma}_z$-symmetry. The error is evaluated with respect to a reference T-matrix that has been symmetrized to restore $\zsym$ according to (\ref{eq:symmetrize}).}
\label{fig-symmetry_error}
\end{figure}

When calculating the T-matrix of the cylinder example in the main text, we observe a small deviation from the expected xy-plane mirror symmetry, which is attributed to an asymmetry in the numerical discretization. In particular, the mesh breaks this mirror symmetry. In Figure~\ref{fig-symmetry_error}, we quantify the relative error in the T-matrix due to this broken symmetry.

Here we denote the mirror transformation about the xy-plane as $\boldsymbol{\sigma}_z$. For an object that possesses $\zsym$, the T-matrix must fulfill the relation \citep{Asadova_2025}:

\begin{equation}
T_{\ell,m,\ell^\prime,m^\prime}^{ij} = (-1)^{i+j+m+m^\prime+\ell+\ell^\prime} T_{\ell,m,\ell^\prime,m^\prime}^{ij} \, .
\end{equation}

In other words, all T-matrix elements that connect VSWs of different $\boldsymbol{\sigma}_z$-parity must be zero.
The relation directly results from $\mathbf{T} = \boldsymbol{\sigma}_z \mathbf{T} \boldsymbol{\sigma}_z$ and can be enforced by symmetrizing the numerically evaluated T-matrix as:

\begin{equation}
\label{eq:symmetrize}
\mathbf{T}_\mathrm{sym} = \frac{1}{2} \left( \mathbf{T} + \boldsymbol{\sigma}_z \mathbf{T} \boldsymbol{\sigma}_z \right) \, .
\end{equation}

Comparing the relative error shown in Figure~\ref{fig-symmetry_error} to the deviations found in Supplement III (`FEM Convergence Study'), it is clear that the symmetry breaking is minimal. Nonetheless, to increase clarity, the results shown in the main text for the cylinder example have been obtained after applying the symmetrization procedure.

\begin{figure}[!htbp]
\centering
\includegraphics[width=0.95\linewidth]{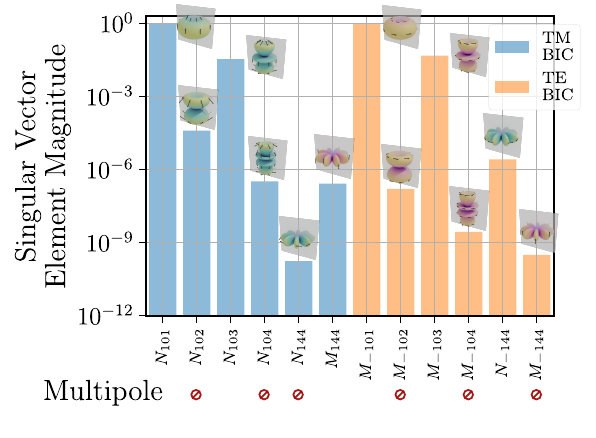}
\caption[]{Multipolar contents of the quasi-dual BICs of the cylinder example without applying the symmetrization procedure to restore $\boldsymbol{\sigma}_z$-symmetry. In contrast to Figure 6 of the main text, additional multipolar contributions appear. Our group theoretical analysis of the permitted multipolar contributions when $\boldsymbol{\sigma}_z$-symmetry is broken is in agreement with the observed additional multipolar contributions. Multipoles not allowed for the $D_{4h}$ symmetric lattice are marked with forbidden symbols below the bars.}
\label{fig-multipoles-broken-symmetry}
\end{figure}

\textbf{Effect of Broken $\boldsymbol{\sigma}_z$-Symmetry on Multipolar Contents of BICs:}
As shown in Figure~\ref{fig-multipoles-broken-symmetry}, if the symmetrization is not applied, the BICs couple to additional multipoles (see Figure 6 of the main text for comparison). To understand the additional multipolar contributions, we have to consider the group theoretical implications of breaking $\boldsymbol{\sigma}_z$-symmetry. The point group of the lattice of $\boldsymbol{\sigma}_z$-symmetric cylinders is $D_{4h}$, while breaking $\boldsymbol{\sigma}_z$-symmetry reduces the point group to $C_{4v}$. To find the permitted multipolar contributions, we will thus have to find which irreducible representations (irreps) the different multipoles correspond to in $C_{4v}$. We will shortcut this analysis by leveraging the known correspondence between irreps and multipoles for $D_{4h}$ shown in Table~\ref{tab-mutl-D4h} \citep{Gladyshev_2020}. By breaking the $\boldsymbol{\sigma}_z$-symmetry, the transformations $C'_{2}$, $C''_{2}$, $i$, $S_{4}$ and $\sigma_{h}$  (see Table~\ref{tab-char-D4h} and Table~\ref{tab-char-C4v}) are dropped. As a consequence, the irreps of $D_{4h}$ merge into the irreps of $C_{4v}$ according to the irreps correlation diagram in Table~\ref{tab-merging-D4h-C4v} \citep{wilson1955molecular}. Thus, we can directly find the correspondence between multipoles and irreps of $C_{4v}$ by merging the multipoles accordingly.

\begin{table*}
\centering
\caption[]{Character Table of $D_{4h}$ \citep{cotton1991chemical}.}
\label{tab-char-D4h}
\begin{tabular}{p{\dimexpr 0.091\linewidth-2\tabcolsep}p{\dimexpr 0.091\linewidth-2\tabcolsep}p{\dimexpr 0.091\linewidth-2\tabcolsep}p{\dimexpr 0.091\linewidth-2\tabcolsep}p{\dimexpr 0.091\linewidth-2\tabcolsep}p{\dimexpr 0.091\linewidth-2\tabcolsep}p{\dimexpr 0.091\linewidth-2\tabcolsep}p{\dimexpr 0.091\linewidth-2\tabcolsep}p{\dimexpr 0.091\linewidth-2\tabcolsep}p{\dimexpr 0.091\linewidth-2\tabcolsep}p{\dimexpr 0.091\linewidth-2\tabcolsep}}
\toprule
$D_{4h}$ & $E$ & $2C_{4}(z)$ & $C_{2}$ & $2C'_{2}$ & $2C''_{2}$ & $i$ & $2S_{4}$ & $\sigma_{h}$ & $2\sigma_{v}$ & $2\sigma_{d}$ \\
\hline
$A_{1g}$ & $1$ & $+1$ & $+1$ & $+1$ & $+1$ & $+1$ & $+1$ & $+1$ & $+1$ & $+1$ \\
$A_{2g}$ & $1$ & $+1$ & $+1$ & $-1$ & $-1$ & $+1$ & $+1$ & $+1$ & $-1$ & $-1$ \\
$B_{1g}$ & $1$ & $-1$ & $+1$ & $+1$ & $-1$ & $+1$ & $-1$ & $+1$ & $+1$ & $-1$ \\
$B_{2g}$ & $1$ & $-1$ & $+1$ & $-1$ & $+1$ & $+1$ & $-1$ & $+1$ & $-1$ & $+1$ \\
$E_{g}$ & $2$ & $\zz$ & $-2$ & $\zz$ & $\zz$ & $+2$ & $\zz$ & $-2$ & $\zz$ & $\zz$ \\
$A_{1u}$ & $1$ & $+1$ & $+1$ & $+1$ & $+1$ & $-1$ & $-1$ & $-1$ & $-1$ & $-1$ \\
$A_{2u}$ & $1$ & $+1$ & $+1$ & $-1$ & $-1$ & $-1$ & $-1$ & $-1$ & $+1$ & $+1$ \\
$B_{1u}$ & $1$ & $-1$ & $+1$ & $+1$ & $-1$ & $-1$ & $+1$ & $-1$ & $-1$ & $+1$ \\
$B_{2u}$ & $1$ & $-1$ & $+1$ & $-1$ & $+1$ & $-1$ & $+1$ & $-1$ & $+1$ & $-1$ \\
$E_{u}$ & $2$ & $\zz$ & $-2$ & $\zz$ & $\zz$ & $-2$ & $\zz$ & $+2$ & $\zz$ & $\zz$ \\
\bottomrule
\end{tabular}
\end{table*}

\begin{table*}
\centering
\caption[]{Multipoles grouped by irreducible representations of $D_{4h}$ according to which they transform. Here $s,k \in \mathbb{N}^+$.}
\label{tab-mutl-D4h}
\begin{tabular}{p{\dimexpr 0.250\linewidth-2\tabcolsep}p{\dimexpr 0.250\linewidth-2\tabcolsep}p{\dimexpr 0.250\linewidth-2\tabcolsep}p{\dimexpr 0.250\linewidth-2\tabcolsep}}
\toprule
Parity & Irrep $D_{4h}$, (dim) & Electric & Magnetic \\
\hline
Odd & $A_{1u}$, (1) & $N_{ -1,4s,2k -1}$ & $M_{ -1,0,2k}$, $M_{ -1,4s,2k}$ \\
Odd & $A_{2u}$, (1) & $N_{1,0,2k -1}$, $N_{1,4s,2k -1}$ & $M_{1,4s,2k}$ \\
Odd & $B_{1u}$, (1) & $N_{ -1,4s -2,2k -1}$ & $M_{ -1,4s -2,2k}$ \\
Odd & $B_{2u}$, (1) & $N_{1,4s -2,2k -1}$ & $M_{1,4s -2,2k}$ \\
Odd & $E_{g}$, (2) & $N_{ -1,4s -3,2k}$, $N_{ -1,4s -1,2k}$ & $M_{ -1,4s -3,2k -1}$, $M_{ -1,4s -1,2k -1}$ \\
 &  & $N_{1,4s -3,2k}$, $N_{1,4s -1,2k}$ & $M_{1,4s -3,2k -1}$, $M_{1,4s -1,2k -1}$ \\
 &  &  &  \\
Even & $A_{1g}$, (1) & $N_{1,0,2k}$, $N_{1,4s,2k}$ & $M_{1,4s,2k -1}$ \\
Even & $A_{2g}$, (1) & $N_{ -1,4s,2k}$ & $M_{ -1,0,2k -1}$, $M_{ -1,4s,2k -1}$ \\
Even & $B_{1g}$, (1) & $N_{1,4s -2,2k}$ & $M_{1,4s -2,2k -1}$ \\
Even & $B_{2g}$, (1) & $N_{ -1,4s -2,2k}$ & $M_{ -1,4s -2,2k -1}$ \\
Even & $E_{u}$, (2) & $N_{ -1,4s -3,2k -1}$, $N_{ -1,4s -1,2k -1}$ & $M_{ -1,4s -3,2k}$, $M_{ -1,4s -1,2k}$ \\
 &  & $N_{1,4s -3,2k -1}$, $N_{1,4s -1,2k -1}$ & $M_{1,4s -3,2k}$, $M_{1,4s -1,2k}$ \\
\bottomrule
\end{tabular}
\end{table*}

\begin{table}
\centering
\caption[]{Irreps correlation diagram for $D_{4h}$ and $C_{4v}$ \citep{wilson1955molecular}.}
\label{tab-merging-D4h-C4v}
\begin{tabular}{p{\dimexpr 0.333\linewidth-2\tabcolsep}p{\dimexpr 0.333\linewidth-2\tabcolsep}p{\dimexpr 0.333\linewidth-2\tabcolsep}}
\toprule
$C_{4v}$ & $\leftarrow$ & $D_{4h}$ \\
\hline
$A_1$ & $\leftarrow$ & $A_{1g}$, $A_{2u}$ \\
$A_2$ & $\leftarrow$ & $A_{2g}$, $A_{1u}$ \\
$B_1$ & $\leftarrow$ & $B_{1g}$, $B_{2u}$ \\
$B_2$ & $\leftarrow$ & $B_{2g}$, $B_{1u}$ \\
$E$ & $\leftarrow$ & $E_{g}$, $E_{u}$ \\
\bottomrule
\end{tabular}
\end{table}

\begin{table}
\centering
\caption[]{Character Table of $C_{4v}$ \citep{cotton1991chemical}.}
\label{tab-char-C4v}
\begin{tabular}{p{\dimexpr 0.167\linewidth-2\tabcolsep}p{\dimexpr 0.167\linewidth-2\tabcolsep}p{\dimexpr 0.167\linewidth-2\tabcolsep}p{\dimexpr 0.167\linewidth-2\tabcolsep}p{\dimexpr 0.167\linewidth-2\tabcolsep}p{\dimexpr 0.167\linewidth-2\tabcolsep}}
\toprule
$C_{4v}$ & $E$ & $2C_{4}(z)$ & $C_{2}$ & $2\sigma_{v}$ & $2\sigma_{d}$ \\
\hline
$A_{1}$ & $+1$ & $+1$ & $+1$ & $+1$ & $+1$ \\
$A_{2}$ & $+1$ & $+1$ & $+1$ & $-1$ & $-1$ \\
$B_{1}$ & $+1$ & $-1$ & $+1$ & $+1$ & $-1$ \\
$B_{2}$ & $+1$ & $-1$ & $+1$ & $-1$ & $+1$ \\
$E$ & $+2$ & $\zz$ & $-2$ & $\zz$ & $\zz$ \\
\bottomrule
\end{tabular}
\end{table}

\begin{table}
\centering
\caption[]{Multipoles grouped by irreducible representations of $C_{4v}$ according to which they transform. Here $\kappa \in \mathbb{N}^+$. We introduce $\kappa$ in contrast to $k$ to highlight the more permissive selection rules as compared to the $D_{4h}$ case.}
\label{tab-mult-C4v}
\begin{tabular}{p{\dimexpr 0.333\linewidth-2\tabcolsep}p{\dimexpr 0.333\linewidth-2\tabcolsep}p{\dimexpr 0.333\linewidth-2\tabcolsep}}
\toprule
Irrep $C_{4v}$, (dim) & Electric & Magnetic \\
\hline
$A_{1}$, (1) & $N_{1,0,\kappa}$, $N_{1,4s,\kappa}$ & $M_{1,4s,\kappa}$ \\
$A_{2}$, (1) & $N_{ -1,4s,\kappa}$ & $M_{ -1,0,\kappa}$, $M_{ -1,4s,\kappa}$ \\
$B_{1}$, (1) & $N_{1,4s -2,\kappa}$ & $M_{1,4s -2,\kappa}$ \\
$B_{2}$, (1) & $N_{ -1,4s -2,\kappa}$ & $M_{ -1,4s -2,\kappa}$ \\
$E$, (2) & $N_{ -1,4s -3,\kappa}$, $N_{ -1,4s -1,\kappa}$ & $M_{ -1,4s -3,\kappa}$, $M_{ -1,4s -1,\kappa}$ \\
 & $N_{1,4s -3,\kappa}$, $N_{1,4s -1,\kappa}$ & $M_{1,4s -3,\kappa}$, $M_{1,4s -1,\kappa}$ \\
\bottomrule
\end{tabular}
\end{table}

The resulting correspondence is shown in Table~\ref{tab-mult-C4v}. In a final step, to find the multipoles that can contribute to the quasi dual BICs, we find the irrep according to which the electric and magnetic z-oriented dipoles ($N_{101}$ and $M_{ -101}$) transform in $C_{4v}$, which are $A_1$ and $A_2$, respectively. The multipoles observed in Figure~\ref{fig-multipoles-broken-symmetry} are just the multipoles that transform according to these irreps. The multipoles that are forbidden for $D_{4h}$ symmetry, are marked with forbidden symbols below the bars in Figure~\ref{fig-multipoles-broken-symmetry}. Compared to allowed multipoles of comparable degree these have significantly smaller contributions, which is consistent with the small symmetry breaking due to the discretization. However, the analysis also reveils, that the spurious multipolar contributions are on the degree of and even stronger than some of the allowed multipolar contributions of higher degree. That is why we have applied the symmetrization procedure to restore $\zsym$ in the main text for clarity.

\null
\clearpage
\begin{figure*}[!htbp]
\centering
\includegraphics[width=0.9\linewidth]{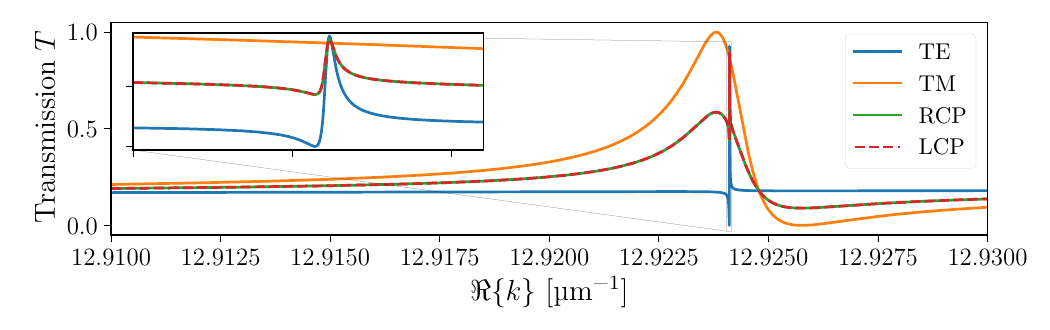}
\caption[]{Transmission spectrum of the lattice of cylinders considered in the main text under linearly and circularly polarized plane wave illumination with an in-plane wavevector component of $k_x\,\mathord{=}\,$\qty{0.1}{\per\micro\meter}. This figure should be understood as a zoom-in to Figure 4 of the main text. The response to additional excitations, in addition to the previously shown TM-polarized illumination, was added. Outside the chosen wavenumber range, the response is largely unaffected by the chosen polarization, as the deviation from normal incidence is small.}
\label{fig-circ_bic}
\end{figure*}

\begin{figure*}[!htbp]
\centering
\includegraphics[width=0.9\linewidth]{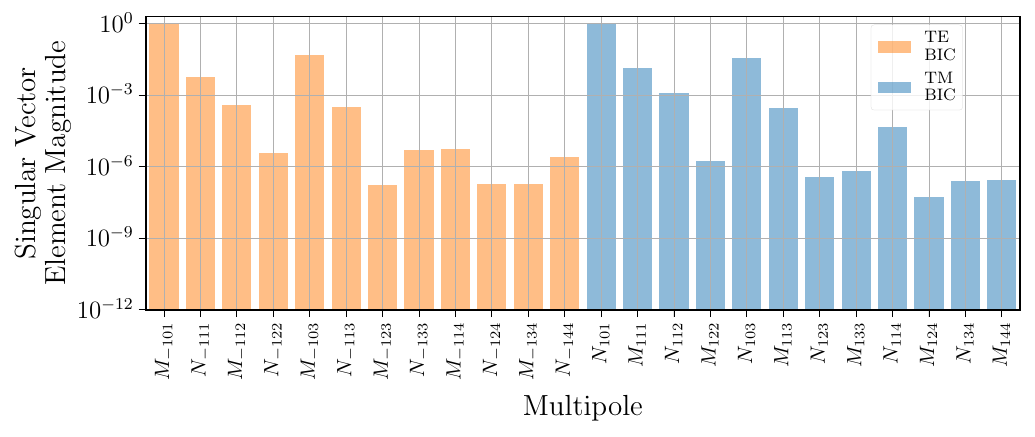}
\caption[]{Multipolar contents of the quasi-dual qBICs. As in Figure~\ref{fig-circ_bic} we choose an in-plane wavevector component of $k_x\,\mathord{=}\,$\qty{0.1}{\per\micro\meter}. The multipolar contents are evaluated as described in the main text.}
\label{fig-multipoles-offgamma}
\end{figure*}

\section{Off-Axis Illumination}

In the main text, we have focused on the case of a normal-incidence illumination of the metasurface, corresponding to an excitation at the $\Gamma$-point in the reciprocal lattice. In this section, we briefly discuss the effects of off-$\Gamma$ illumination, which can be treated straightforwardly within our framework by considering oblique incidence plane waves. Here we choose to move along the high-symmetry line from $\Gamma$ to $X$ by introducing a small in-plane wavevector component $k_x\,\mathord{=}\,$\qty{0.1}{\per\micro\meter}.

\subsection{Plane Wave Illumination}

In Figure 4 of the main text, we show the transmission spectrum of the lattice of cylinders treated in the main text under near-normal plane wave illumination. Here, we show the response to TM- and TE-polarized linear plane waves as well as to right- and left-circularly polarized plane waves. While the off-resonant response does not depend significantly on the chosen polarization (at normal incidence, all lineraly polarized waves are scattered with the same complex amplitudes due to the fourfold symmetry), the response in the vicinity of the qBICs is polarization dependent. In particular, the two qBICs can be separately addressed by the different polarizations: The E-field of TM and TE plane waves as even and odd symmetry with respect to the $xz$-plane, respectively, which is reflected in the symmetries of the two qBICs. The qBICs are entirely independent of each other. They coincide because we have chosen the lattice constant appropriately to leverage the meta-atoms' directional duality. Due to the slight incidence angle, the resonances are no longer decoupled from the far field. Their radiation is mediated by the in-plane dipole moments and other higher-degree multipoles.

\vfill\eject
In contrast to the z-oriented dipoles, the meta-atom is not dual with respect to these other multipoles. This difference manifests itself in differing linewidths of the two qBICs, as their coupling to the far field differs. These unequal linewidths lead to a peculiar nesting of spectral features under circularly polarized illumination (see green and red-dotted lines in Figure~\ref{fig-circ_bic}). Keep in mind that the features shown in Figure~\ref{fig-circ_bic} are already narrow Fano-like resonances superimposed on the in comparison slowly varying resonant background shown in Figure 4 of the main text. Given the already narrow feature of the TM-qBIC, the TE-qBIC appears as an even narrower feature superimposed on top, together forming a spectral feature that bridges approximately four orders of magnitude in spectral resolution from the broadest to the narrowest feature.

\subsection{Multipolar Contents of the qBICs}

For completeness, we show in Figure~\ref{fig-multipoles-offgamma} the multipolar contents of the quasi-dual qBICs (where the symmetry protection is broken by a near-normal illumnination). Once again, these results agree nicely with the group-theoretical considerations of \cite{sadrieva_multipolar_2019}: The TE-qBIC contains only multipoles that transform according to $B_2$, while the TM-qBIC contains only multipoles that transform according to $B_1$.

\end{document}